\documentclass[11pt]{article}
\DeclareMathAlphabet{\scr}{U}{rsfs}{m}{n}

\pdfoutput=1
\usepackage{latexsym}
\usepackage{epsfig}
\usepackage[mathscr]{eucal}
\usepackage{amsfonts}
\usepackage{amscd}
\usepackage{cite}
\usepackage{array}
\usepackage{amssymb}
\usepackage{colordvi}
\usepackage[centertags]{amsmath}
\usepackage{enumerate}
\usepackage{graphicx}
\usepackage{booktabs}
\usepackage{theorem}
\usepackage[footnotesize]{caption}
\usepackage{soul}
\usepackage{url}
\usepackage{mcite}
\usepackage{slashed}
\usepackage{color}
\usepackage{ulem}
\usepackage{bbm}
\usepackage[utf8]{inputenc}
\newcommand{\newc}{\newcommand}
\newc{\be}{\begin{equation}}
\newc{\ee}{\end{equation}}
\newc{\bea}{\begin{eqnarray}}
\newc{\eea}{\end{eqnarray}}
\newc{\ol}{\overline}
\newc{\wt}{\widetilde}
\newc{\bs}{\boldsymbol}
\newc{\m}{\mathcal}
\newc{\la}{\langle}
\newc{\ra}{\rangle}

\newcommand{\beq}{\begin{eqnarray}}
\newcommand{\eeq}{\end{eqnarray}}
\newcommand{\bpmatrix}{\begin{pmatrix}}
\newcommand{\epmatrix}{\end{pmatrix}}
\newcommand{\ba}{\begin{array}}
\newcommand{\ea}{\end{array}}

\renewcommand{\ol}{\text{1l}}

\renewcommand{\eqref}[1]{Eq.~(\ref{#1})}

\newcommand{\bc}{\begin{center}}
\newcommand{\ec}{\end{center}}

\newcommand{\gsim}{\raisebox{-0.13cm}{~\shortstack{$>$ \\[-0.07cm]
      $\sim$}}~}
\newcommand{\lsim}{\raisebox{-0.13cm}{~\shortstack{$<$ \\[-0.07cm]
      $\sim$}}~}
\newcommand{\s}{\newline \vspace*{-3.5mm}}

\begin{document}

\title{
\vspace*{-3.7cm}
\phantom{h} \hfill\mbox{\small KA-TP-10-2017}\\[-1.1cm]
\phantom{h} \hfill\mbox{\small DESY 17-046}
\\[1cm]
\textbf{Phenomenological Comparison of Models with Extended Higgs Sectors \\[4mm]}}

\date{}
\author{
Margarete M\"{u}hlleitner$^{1\,}$\footnote{E-mail:
\texttt{milada.muehlleitner@kit.edu}} ,
Marco O. P. Sampaio$^{2\,}$\footnote{E-mail:
\texttt{msampaio@ua.pt}} ,
Rui Santos$^{3,4,5\,}$\footnote{E-mail:
  \texttt{rasantos@fc.ul.pt}} ,
Jonas Wittbrodt$^{1,6\,}$\footnote{E-mail: \texttt{jonas.wittbrodt@desy.de}}
\\[5mm]
{\small\it
$^1$Institute for Theoretical Physics, Karlsruhe Institute of Technology,} \\
{\small\it 76128 Karlsruhe, Germany}\\[3mm]
{\small\it
$^2$Departamento de F\'{\i}sica, Universidade de Aveiro and CIDMA,} \\
{\small\it Campus de Santiago, 3810-183 Aveiro, Portugal}\\[3mm]
{\small\it
$^3$ISEL -
 Instituto Superior de Engenharia de Lisboa,} \\
{\small \it   Instituto Polit\'ecnico de Lisboa
 1959-007 Lisboa, Portugal} \\[3mm]
{\small\it
$^4$Centro de F\'{\i}sica Te\'{o}rica e Computacional,
    Faculdade de Ci\^{e}ncias,} \\
{\small \it    Universidade de Lisboa, Campo Grande, Edif\'{\i}cio C8
  1749-016 Lisboa, Portugal} \\[3mm]
{\small\it
$^5$LIP, Departamento de F\'{\i}sica, Universidade do Minho, 4710-057 Braga, Portugal}\\[3mm]
{\small\it
$^6$Deutsches Elektronen-Synchrotron DESY, Notkestra{\ss}e 85, 22607
Hamburg, Germany}
}

\maketitle

\begin{abstract}
Beyond the Standard Model (SM) extensions usually include
extended Higgs sectors. Models with singlet or doublet
fields are the simplest ones that are compatible with the $\rho$
parameter constraint. The discovery of new non-SM Higgs
bosons and the identification of the
underlying model requires dedicated Higgs properties analyses. In this paper, we compare several Higgs sectors
featuring 3 CP-even neutral Higgs bosons that are also motivated by
their simplicity and their capability to solve some of the flaws of
the SM. They are: the SM extended by a complex singlet field (CxSM),
the singlet extension of the 2-Higgs-Doublet Model (N2HDM), and
the Next-to-Minimal Supersymmetric SM extension (NMSSM). In
addition, we analyse the CP-violating 2-Higgs-Doublet Model (C2HDM), which
provides 3 neutral Higgs bosons with a pseudoscalar admixture.
This allows us to compare the effects of singlet and pseudoscalar admixtures.
Through dedicated scans of the allowed parameter space of
the models, we analyse
the phenomenologically viable scenarios from the view point of the SM-like Higgs boson and of the signal rates of the non-SM-like Higgs bosons to be found. In particular, we analyse the effect of singlet/pseudoscalar admixture, and the potential to differentiate these models in the near future.
This is supported by a study of couplings sums of the
Higgs bosons to massive gauge bosons and to fermions, where we identify features that allow us to distinguish the models, in particular when only part of the Higgs spectrum is discovered. Our
results can be taken as guidelines for
future LHC data analyses, by the ATLAS and CMS experiments, to identify specific benchmark points aimed at revealing the underlying model.
\end{abstract}
\thispagestyle{empty}
\vfill
\newpage
\setcounter{page}{1}

\section{Introduction}
While the discovery of the Higgs boson by the LHC experiments ATLAS \cite{Aad:2012tfa}
and CMS \cite{Chatrchyan:2012ufa} has been a great success for
particle physics and the Standard Model (SM) in particular, the
unsolved puzzles of the SM call for New Physics (NP) extensions beyond
the SM (BSM). Since we are still lacking any direct
discovery of BSM physics, the Higgs sector itself has
become a tool in the search for NP. The latter can manifest
itself in various ways \cite{Englert:2014uua}. The discovery of
additional Higgs bosons or the measurement of new sources of CP
violation in the Higgs sector would be a direct observation of BSM
physics. Indirect hints would be given by deviations of the Higgs
couplings from the SM expectations. As the discovered Higgs boson
behaves very SM-like
\cite{Khachatryan:2014kca,Aad:2015mxa,Khachatryan:2014jba,Aad:2015gba}
the revelation of such deviations requires, on the one hand, very
precise measurements from the
experiments and, on the other hand, very precise predictions from the
theory side. In parallel to the increase in precision, observables
have to be identified that allow for the identification of NP,
in particular the nature of the underlying model. Thus, the pattern of the
coupling deviations gives information on the specific model that may be responsible. Production rates may be exploited to exclude some
of the models or to single out the model realized in nature. In the ideal
case smoking gun signatures are identified that unmask the model
behind NP. \s

The immense amount of possible BSM Higgs sectors
calls for a strategy on the choice of the models to be
investigated. Any NP model has to provide a Higgs boson with a mass of
125.09~GeV \cite{Aad:2015zhl} that behaves SM-like. The model has to
fulfil the exclusion bounds from Higgs and NP searches, the $B$-physics
and various low-energy constraints and to be compatible with the electroweak
precision data. Furthermore, the theoretical constraints on the Higgs potential,
{\it i.e.}~that it is bounded from below, that the chosen vacuum is
the global minimum at tree level and that perturbative unitarity holds, have to be
fulfilled. Among the weakly coupled models those with singlet or
doublet extended Higgs sectors belong to the simplest extensions that
comply with the $\rho$ parameter constraint.
For this class of models we have analysed, in previous works, their distinction based on
collider phenomenology. In \cite{Englert:2014uua} we
studied the implications of precision measurements of the Higgs
couplings for NP scales and showed how coupling sum rules can be used
to tell the Next-to-Minimal Supersymmetric extension (NMSSM) from the
Minimal Supersymmetric extension (MSSM). We reassessed this question in
\cite{King:2014xwa} in the framework of specific NMSSM benchmarks.
In \cite{Fontes:2015xva}, for the 2-Higgs-Doublet Model
(2HDM), and in \cite{King:2015oxa}, for the NMSSM, we demonstrated how
the simultaneous
measurements of Higgs decays involving the 125~GeV Higgs boson, the
$Z$ boson and one additional Higgs boson undoubtedly distinguish a
CP-violating from a
CP-conserving Higgs sector. In \cite{Costa:2015llh} we found that the distinction
of the complex-singlet extended SM (CxSM) from the NMSSM based on
Higgs-to-Higgs decays is only possible through final states with two different scalars.
The authors of \cite{Gupta:2012mi,Gupta:2013zza} attacked the task of
differentiating NP at the LHC from a different perspective by
asking how well the Higgs mass and couplings need to be measured to
see deviations from the SM. In a similar spirit we investigated in
\cite{Grober:2016wmf} if NP could first be seen in Higgs pair
production taking into account Higgs coupling constraints. \s

In this work we elaborate further on the distinction of NP models
based on LHC collider phenomenology. We go beyond our previous works
by comparing a larger class of models that are, to some extent, similar
in their Higgs sector but involve different symmetries. We explore how
this manifests in the Higgs phenomenology and how it can be exploited to
differentiate the models. With the guiding principle that the models
are able to solve some of the questions of the SM while remaining
compatible with the given constraints, we investigate in this work the
simplest extensions featuring 3 neutral CP-even Higgs bosons. This
particular scenario is phenomenologically interesting because it
allows for Higgs-to-Higgs decays into final states with {\it two
  different} Higgs bosons that lead to rather high
rates, see {\it e.g.}~\cite{Costa:2015llh}.
At the same time we go beyond
the largely studied minimal versions with 2 neutral CP-even Higgs
bosons, the 2HDM and the MSSM. We will investigate the CxSM (the SM
extended by a complex singlet field) in its broken phase,
\cite{Coimbra:2013qq,Costa:2015llh}, the Next-to-Minimal 2HDM (N2HDM, the 2HDM \cite{Gunion:1989we,Lee:1973iz,Branco:2011iw}
extended by a singlet field)
\cite{He:2008qm,Grzadkowski:2009iz,Logan:2010nw,Boucenna:2011hy,He:2011gc,Bai:2012nv,He:2013suk,Cai:2013zga,Wang:2014elb,Drozd:2014yla,Campbell:2015fra,vonBuddenbrock:2016rmr,Chen:2013jvg,Muhlleitner:2016mzt},
and, as representative for a supersymmetric (SUSY) model, the NMSSM
\cite{Fayet:1974pd,Barbieri:1982eh,Dine:1981rt,Nilles:1982dy,Frere:1983ag,Derendinger:1983bz,Ellis:1988er,Drees:1988fc,Ellwanger:1993xa,Ellwanger:1995ru,Ellwanger:1996gw,Elliott:1994ht,King:1995vk,Franke:1995tc,Maniatis:2009re,Ellwanger:2009dp}. While in all
three models the singlet admixture to the Higgs mass eigenstates
decreases their couplings to the SM particles, they are considerably
different: The NMSSM is subject to SUSY
relations to be fulfilled while the CxSM is much simpler than
the N2HDM and NMSSM, which contain a charged Higgs boson and,
additionally, one and two CP-odd Higgs bosons, respectively.
We will also compare the phenomenological effects of coupling modifications through singlet admixture with the corresponding effects caused by CP violation.
For this purpose we include the complex 2HDM (C2HDM)
\cite{Ginzburg:2002wt,Khater:2003wq,ElKaffas:2006gdt,ElKaffas:2007rq,WahabElKaffas:2007xd,Osland:2008aw,Grzadkowski:2009iz,Arhrib:2010ju,Barroso:2012wz,Fontes:2014xva} in our study.\footnote{We
  do not include the possibility of a CP-violating N2HDM or NMSSM as our
focus here is on the comparison of Higgs sectors with 3 neutral Higgs
bosons that either have a singlet or a CP-admixture, whereas those
models would increase the number of CP-violating Higgs bosons beyond
3.} In this model all 3 neutral Higgs bosons mix to form CP-violating mass eigenstates,
in contrast to the real 2HDM which features 2 CP-even and 1
CP-odd Higgs boson. The measurement of CP violation is experimentally
very challenging and, at a first stage, in the discovery of the neutral
Higgs bosons of the C2HDM they can be misidentified as CP-even or CP-odd Higgs
bosons. In such an experimental scenario we have a clear connection to
our other CP-conserving models that also contain three neutral Higgs
bosons mixing. This  allows us to compare the effect of the singlet
admixtures with the effect of CP violation on the Higgs couplings and associated physical processes. These two different ways of achieving coupling modifications may induce a considerably different Higgs phenomenology that might then be revealed
by the appropriate observables. Finally, all these models may
solve problems of the SM. Depending on the model and (possibly) on its spontaneous symmetry breaking phase it may
{\it e.g.}~provide a Dark Matter (DM) candidate, lead to
successful baryogenesis, weaken the hierarchy problem or solve the $\mu$
problem of the MSSM \cite{Gunion:1989we,Martin:1997ns,Dawson:1997tz,Djouadi:2005gj}. \s

For all investigated models we will perform parameter scans by taking into
account the experimental and theoretical constraints. We will
investigate the mass distributions and the properties of
the 125 GeV Higgs boson as a function of the singlet admixture and of
the pseudoscalar admixture, respectively. We will study the production
rates of the non-SM Higgs bosons and investigate Higgs coupling
sums. We aim to answer the following
questions: To which extent can LHC Higgs
phenomenology, in particular signal rates and coupling
measurements, be exploited to distinguish between these models with extended
Higgs sectors? Are we able to disentangle the models based on Higgs
rate measurements? Can the pattern of the couplings of the discovered Higgs bosons
point towards possibly missing Higgs bosons in case not all of them have
been discovered? Is it even possible to use coupling sums to
reveal the underlying model? Can the investigation of the couplings
give hints on the underlying NP scale?
With our findings we hope to encourage the experiments to
conduct specific phenomenological analyses and investigate the
relevant observables. We aim to contribute to the endeavour of
revealing the underlying NP model (if realized in
nature) by using all the available data from the LHC experiments.  \s

The outline of the paper is as follows. In section \ref{sec:model} we
will present our models and introduce our notation. Section
\ref{sec:scans} describes the scans with the applied
constraints. With section \ref{sec:pheno} we start our
phenomenological analysis. After presenting the mass distribution of the Higgs
spectra of our models, the phenomenology of the SM-like Higgs boson
will be described. In Sect.~\ref{sec:sigrates} the signal rates of the
non-SM-like Higgs bosons will be presented and discussed. Section
\ref{sec:sumrules} is dedicated to the investigation of the couplings
sums. Our conclusions are given in Sect.~\ref{sec:concl}.

\section{Description of the Models \label{sec:model}}
\setcounter{equation}{0}
In this section we describe the models that we investigate. We start
with the simplest one, the SM extended by a complex singlet field, the
CxSM. We then move on in complexity with the (C)2HDM, the N2HDM and the
NMSSM. We use this description also to set our notation.

\subsection{The Complex Singlet Extension of the SM \label{sec:cxsm}}
In the CxSM a complex singlet field
\beq
\mathbb{S} =  S + iA
\eeq
with hypercharge zero is added to the SM Lagrangian. We study the CxSM since the simpler
extension by a real singlet field, the RxSM, features only two Higgs
bosons. The scalar
potential with a softly broken global $U(1)$ symmetry is given by
\beq
V = \frac{m^2}{2} H^\dagger H + \frac{\lambda}{4} (H^\dagger
H)^2+\frac{\delta_2}{2} H^\dagger H |\mathbb{S}|^2 +
\frac{b_2}{2}|\mathbb{S}|^2+ \frac{d_2}{4} |\mathbb{S}|^4 +
\left(\frac{b_1}{4} \mathbb{S}^2 + a_1 \mathbb{S} +c.c. \right)
\, ,  \label{eq:VCxSM}
\eeq
with the soft-breaking terms in parenthesis. The doublet and complex
singlet fields can be written as
\begin{equation}
H=\dfrac{1}{\sqrt{2}}\left(\begin{array}{c} G^+ \\
    v+h+iG^0\end{array}\right) \quad \mbox{and} \quad
\mathbb{S}=\dfrac{1}{\sqrt{2}}\left[v_S+s+i(v_A+ a)\right] \;,
\label{eq:fieldsCxSM}
\end{equation}
where $v\approx 246$~GeV denotes the vacuum
expectation value (VEV) of the SM Higgs boson $h$ and $v_S$ and $v_A$
are the real and imaginary parts of the complex singlet field VEV,
respectively. We impose a $\mathbb{Z}_2$ symmetry on $A$,
which is equivalent to a symmetry under $\mathbb{S} \to \mathbb{S}^*$. This
forces $a_1$ and $b_1$ to be real. The remaining parameters $m,
\lambda, \delta_2 , b_2$ and $d_2$ are required to be real by
hermiticity of the potential. There are two possible phases consistent
with electroweak symmetry breaking (EWSB) \cite{Coimbra:2013qq}. The
symmetric (or DM) phase, with
$v_A=0$ and $v_S \ne 0$, features only two mixed states plus one DM
candidate, so we focus instead on the broken phase. In the latter all VEVs are
non-vanishing and all three scalars mix with each
other. Introducing the notation $\rho_1 \equiv h$, $\rho_2 \equiv s$ and $\rho_3
\equiv a$, their mass matrix is obtained from the potential in the
physical minimum through ($i,j=1,2,3$)
\beq
({\cal M}^2)_{ij} = \left\langle \frac{\partial^2 V}{\partial \rho_i
  \partial \rho_j} \right\rangle \;,
\eeq
where the brackets denote the vacuum.
The three mass eigenstates $H_i$ are obtained from the
gauge eigenstates $\rho_i$ by means of the rotation matrix $R$
as
\beq
\left( \begin{array}{c} H_1\\ H_2\\ H_3 \end{array} \right) = R
\left( \begin{array}{c} \rho_1 \\ \rho_2 \\ \rho_3 \end{array} \right) \, ,
\label{eq:rotsinglet}
\eeq
with
\beq
R\, {\cal M}^2\, R^T = \textrm{diag} \left(m_{H_1}^2, m_{H_2}^2, m_{H_3}^2 \right),
\eeq
and $m_{H_1} \leq m_{H_2} \leq m_{H_3}$ denoting the masses of the neutral Higgs
bosons. Introducing the abbreviations $s_{i} \equiv \sin \alpha_i$ and
$c_{i} \equiv \cos \alpha_i$ with
\beq
-\frac{\pi}{2} \le \alpha_i < \frac{\pi}{2} \,,
\label{eg:alpharanges}
\eeq
the mixing matrix $R$ can be parametrized as
\beq
R =\left( \begin{array}{ccc}
c_{1} c_{2} & s_{1} c_{2} & s_{2}\\
-(c_{1} s_{2} s_{3} + s_{1} c_{3})
& c_{1} c_{3} - s_{1} s_{2} s_{3}
& c_{2} s_{3} \\
- c_{1} s_{2} c_{3} + s_{1} s_{3} &
-(c_{1} s_{3} + s_{1} s_{2} c_{3})
& c_{2}  c_{3}
\end{array} \right) \;.
\label{eq:singletmatrix}
\eeq

The model has seven independent parameters, and we choose as input
parameters the set
\beq
\alpha_1 \;, \quad \alpha_2 \;, \quad \alpha_3 \;, \quad v \;, \quad v_S \;,
\quad m_{H_1} \quad \mbox{and} \quad m_{H_3} \;. \label{eq:cxsminput}
\eeq
The VEV $v_A$ and the mass $m_{H_2}$ are dependent parameters. In the scans
that we will perform they are determined internally by the program
{\tt ScannerS} \cite{Coimbra:2013qq,ScannerS} in accordance with the
minimum conditions of the vacuum. \s

The couplings $\lambda_i^{(p)}$ of the Higgs mass eigenstates $H_i$ to
the SM particles, denoted by $p$, are all modified by the same
factor. In terms of the couplings $\lambda_{h_{\text{SM}}}^{(p)}$ of
the SM Higgs boson $h_{\text{SM}}$ they read
\beq
\lambda_i^{(p)} = R_{i1} \lambda_{h_{\text{SM}}}^{(p)} \;.
\eeq
The trilinear Higgs self-couplings are obtained from the terms cubic
in the fields in the potential $V$ of
Eq.~(\ref{eq:VCxSM}) after expanding the doublet and singlet fields
about their VEVs and rotating to the mass eigenstates. Their explicit
expressions together with the quartic couplings can be found in
appendix B.1 of \cite{Costa:2015llh}. If kinematically allowed, the trilinear Higgs couplings induce Higgs-to-Higgs decays that change the
total widths of the $H_i$ and hence their branching ratios to the SM
particles. The branching ratios including the state-of-the art higher
order QCD corrections and possible off-shell decays can be obtained
from {\tt sHDECAY}\footnote{The program {\tt sHDECAY} can be downloaded
  from the url: \url{http://www.itp.kit.edu/~maggie/sHDECAY}.}
which is based on the implementation of
the CxSM and also the RxSM both in their symmetric and broken phases in
{\tt HDECAY}~\cite{Djouadi:1997yw,Butterworth:2010ym}. A detailed
description of the program can be found in appendix A of \cite{Costa:2015llh}. \s

\subsection{The C2HDM\label{sec:2hdm}}
In terms of two $SU(2)_L$ Higgs doublets $\Phi_1$ and $\Phi_2$ the
Higgs potential of a general 2HDM with a softly broken global discrete
$\mathbb{Z}_2$ symmetry is given by
\beq
V &=& m_{11}^2 |\Phi_1|^2 + m_{22}^2 |\Phi_2|^2 - \left[m_{12}^2 \Phi_1^\dagger
\Phi_2 + h.c.\right] + \frac{\lambda_1}{2} (\Phi_1^\dagger \Phi_1)^2 +
\frac{\lambda_2}{2} (\Phi_2^\dagger \Phi_2)^2 \nonumber \\
&& + \lambda_3
(\Phi_1^\dagger \Phi_1) (\Phi_2^\dagger \Phi_2) + \lambda_4
(\Phi_1^\dagger \Phi_2) (\Phi_2^\dagger \Phi_1) + \left[\frac{\lambda_5}{2}
(\Phi_1^\dagger \Phi_2)^2 + h.c.\right] \;.
\eeq
The required invariance under the $\mathbb{Z}_2$ transformations
$\Phi_1 \to -\Phi_1$ and $\Phi_2 \to \Phi_2$ guarantees the absence of
tree-level Flavour Changing Neutral Currents (FCNC). Hermiticity forces
all parameters to be real except for the soft $\mathbb{Z}_2$
breaking mass parameter $m_{12}^2$ and the quartic coupling
$\lambda_5$. If $\mbox{arg}(m_{12}^2) = \mbox{arg} (\lambda_5)$, their
complex phases can be absorbed by a basis transformation. In that case
we are left with the real or CP-conserving 2HDM\footnote{Assuming both
vacuum expectation values to be real.}
depending on eight real parameters. Otherwise we are in the framework
of the complex or CP-violating 2HDM. The C2HDM depends on ten real
parameters. In the following, we will use the conventions from
\cite{Fontes:2014xva} for the C2HDM. After EWSB  the neutral
components of the Higgs doublets develop VEVs, which are real in the
CP-conserving case. Allowing for CP violation, there could be in principle a complex
phase between the VEVs of the two doublets. This phase can, however,
always be removed by a change of basis \cite{Ginzburg:2002wt} so,
without loss of generality, we set it to zero. Expanding about the real VEVs $v_{1}$ and $v_{2}$ and
expressing each doublet $\Phi_i$ $(i=1,2)$ in terms of the charged complex field
$\phi_i^+$ and the real neutral CP-even and CP-odd fields $\rho_i$ and
$\eta_i$, respectively, we have
\beq
\Phi_1 = \left(
\begin{array}{c}
\phi_1^+ \\
\frac{v_1 + \rho_1 + i \eta_1}{\sqrt{2}}
\end{array}
\right) \qquad \mbox{and} \qquad
\Phi_2 = \left(
\begin{array}{c}
\phi_2^+ \\
\frac{v_2 + \rho_2 + i \eta_2}{\sqrt{2}}
\end{array}
\right) \;. \label{eq:2hdmdoubletexpansion}
\eeq
The requirement that the minimum of the potential is given by
\beq
\langle \Phi_i \rangle = \left( \begin{array}{c} 0 \\
    \frac{v_i}{\sqrt{2}} \end{array} \right) \label{eq:normalmin}
\eeq
leads to the minimum conditions
\beq
m_{11}^2 v_1 + \frac{\lambda_1}{2} v_1^3 + \frac{\lambda_{345}}{2} v_1
v_2^2 &=& m_{12}^2 v_2 \label{eq:mincond1} \\
m_{22}^2 v_2 + \frac{\lambda_2}{2} v_2^3 + \frac{\lambda_{345}}{2} v_1^2
v_2 &=& m_{12}^2 v_1 \label{eq:mincond2} \\
2\, \mbox{Im} (m_{12}^2) &=& v_1 v_2 \mbox{Im} (\lambda_5)
\;, \label{eq:mincond3}
\eeq
where we have introduced
\beq
\lambda_{345} \equiv \lambda_3 + \lambda_4 + \mbox{Re} (\lambda_5) \;.
\eeq
Using Eqs.~(\ref{eq:mincond1}) and (\ref{eq:mincond2}) we can trade
the parameters $m_{11}^2$ and $m_{22}^2$ for $v_1$ and
$v_2$, while Eq.~(\ref{eq:mincond3}) yields a
relation between the two sources of CP violation in the scalar
potential. This fixes one of the ten parameters of the C2HDM. \s

The Higgs basis \cite{Lavoura:1994fv,Botella:1994cs} $\{ {\cal H}_1,
{\cal H}_2 \}$, in which the second Higgs doublet
${\cal H}_2$ does not get a VEV, is obtained by the rotation
\beq
\left( \begin{array}{c} {\cal H}_1 \\ {\cal H}_2 \end{array} \right) =
R_H \left( \begin{array}{c} \Phi_1 \\
    \Phi_2 \end{array} \right) \equiv
\left( \begin{array}{cc} c_\beta & s_\beta \\ - s_\beta &
    c_\beta \end{array} \right) \left( \begin{array}{c} \Phi_1 \\
    \Phi_2 \end{array} \right) \;,
\eeq
with
\beq
t_\beta \equiv \frac{v_2}{v_1} \;,
\eeq
so that we have
\beq
{\cal H}_1 = \left( \begin{array}{c} G^\pm \\ \frac{1}{\sqrt{2}} (v + H^0
    + i G^0) \end{array} \right) \quad \mbox{and} \qquad
{\cal H}_2 = \left( \begin{array}{c} H^\pm \\ \frac{1}{\sqrt{2}} (R_2
    + i I_2) \end{array} \right) \;.
\eeq
The SM VEV
\beq
v = \sqrt{v_1^2 + v_2^2} \;,
\eeq
along with the massless charged and neutral would-be Goldstone bosons
$G^\pm$ and $G^0$ is now in doublet one, while the charged Higgs mass
eigenstates $H^\pm$ are contained in doublet two. The neutral Higgs
mass eigenstates $H_i$ ($i=1,2,3$) are obtained from the neutral
components of the C2HDM basis, $\rho_1$, $\rho_2$ and $\rho_3 \equiv I_2$,
through the rotation
\beq
\left( \begin{array}{c} H_1 \\ H_2 \\ H_3 \end{array} \right) = R
\left( \begin{array}{c} \rho_1 \\ \rho_2 \\ \rho_3 \end{array} \right)
\;.
\label{eq:c2hdmrot}
\eeq
The orthogonal matrix $R$ diagonalizes the neutral mass matrix
\beq
({\cal M}^2)_{ij} = \left\langle \frac{\partial^2 V}{\partial \rho_i
  \partial \rho_j} \right\rangle \;,
\label{eq:c2hdmmassmat}
\eeq
through
\beq
R {\cal M}^2 R^T = \mbox{diag} (m_{H_1}^2, m_{H_2}^2, m_{H_3}^2) \;.
\eeq
The Higgs bosons are ordered by ascending mass according to $m_{H_1}
\le m_{H_2} \le m_{H_3}$. For the matrix $R$ we choose the same
parametrization as in Eq.~(\ref{eq:singletmatrix}) and the same range as in Eq.~(\ref{eg:alpharanges})
for the mixing angles.
Note that the mass basis and the Higgs basis are related through
\beq
\left( \begin{array}{c} H_1 \\ H_2 \\ H_3 \end{array} \right) = R \widetilde{R}_H
\left( \begin{array}{c} H^0 \\ R_2 \\ I_2 \end{array} \right) \;,
\eeq
with
\beq
\widetilde{R}_H = \left( \begin{array}{cc} R_H^T & 0 \\ 0 & 1 \end{array}
\right) \;.
\eeq

In total, the C2HDM has 9 independent parameters (one was fixed by the
minimisation conditions) that we choose to be
\cite{ElKaffas:2007rq}
\beq
v \approx 246\mbox{ GeV} \;, \quad t_\beta \;, \quad \alpha_{1,2,3}
\;, \quad m_{H_i} \;, \quad m_{H_j} \;, \quad m_{H^\pm} \quad
\mbox{and} \quad m_{12}^2 \;.
\label{eq:2hdminputset}
\eeq
Here $m_{H_i}$ and $m_{H_j}$ denote any two of the three neutral Higgs
bosons. The third mass is dependent and can be obtained from the other
parameters \cite{ElKaffas:2007rq}. For analytic relations between the
set of parameters Eq.~(\ref{eq:2hdminputset}) and the coupling
parameters $\lambda_i$ of the 2HDM Higgs potential, see
\cite{Fontes:2014xva}. \s

The CP-conserving 2HDM is obtained for $\alpha_2 = \alpha_3= 0$ and
$\alpha_1= \alpha + \pi/2$ \cite{Khater:2003wq}. In this case the mass
matrix Eq.~(\ref{eq:c2hdmmassmat}) becomes block diagonal and $\rho_3$ is
the pseudoscalar mass eigenstate $A$, while the CP-even mass
eigenstates $h$ and $H$ are obtained from the gauge eigenstates through the
rotation parametrized in terms of the angle $\alpha$,
\beq
\left( \begin{array}{c} H \\ h \end{array} \right) =
\left( \begin{array}{cc} c_\alpha & s_\alpha \\ -s_\alpha &
    c_\alpha \end{array} \right) \left( \begin{array}{c} \rho_1 \\
    \rho_2 \end{array} \right) \;,
\eeq
with $-\pi/2 \le \alpha < \pi/2$. By convention $m_h \le m_H$. \s

For the computation of the Higgs boson observables entering our
phenomenological analysis we need the couplings of the C2HDM Higgs
bosons. We introduce the Feynman rules for the Higgs couplings $H_i$ to the
massive gauge bosons $V=W,Z$ as
\beq
i \, g_{\mu\nu} \, c(H_i VV) \, g_{H^{\text SM} VV} \;. \label{eq:gaugecoupdef}
\eeq
Here $g_{H^{\text SM} VV}$ denote the SM Higgs coupling factors. In terms of
the gauge boson masses $M_W$ and $M_Z$, the $SU(2)_L$ gauge coupling
$g$ and the Weinberg angle
$\theta_W$ they are given by $g_{H^{\text
    SM} VV} = g M_W$ for $V=W$ and $g M_Z /\cos\theta_W$ for $V=Z$.
With the definition~\eqref{eq:gaugecoupdef} we then have the effective
couplings \cite{Fontes:2014xva}
\beq
c(H_i VV) = c_\beta R_{i1} + s_\beta R_{i2} \;. \label{eq:c2dhmgaugecoup}
\eeq
In order to avoid tree-level FCNCs
one type of fermions is allowed to couple only to one Higgs doublet by
imposing a global $\mathbb{Z}_2$ symmetry under which $\Phi_{1,2} \to
\mp \Phi_{1,2}$. Depending on the $\mathbb{Z}_2$ charge assignments,
there are four phenomenologically different types of 2HDMs summarized
in table~\ref{tab:types}.
\begin{table}
\begin{center}
\begin{tabular}{rccc} \toprule
& $u$-type & $d$-type & leptons \\ \midrule
type I & $\Phi_2$ & $\Phi_2$ & $\Phi_2$ \\
type II & $\Phi_2$ & $\Phi_1$ & $\Phi_1$ \\
lepton-specific & $\Phi_2$ & $\Phi_2$ & $\Phi_1$ \\
flipped & $\Phi_2$ & $\Phi_1$ & $\Phi_2$ \\ \bottomrule
\end{tabular}
\caption{The four Yukawa types of the $\mathbb{Z}_2$-symmetric 2HDM
  defined by the Higgs doublet that couples to each kind of fermions. \label{tab:types}}
\end{center}
\end{table}
The Feynman rules for the Higgs couplings to the fermions can be
derived from the Yukawa Lagrangian
\beq
{\cal L}_Y = - \sum_{i=1}^3 \frac{m_f}{v} \bar{\psi}_f \left[ c^e(H_i
  ff) + i c^o(H_i ff) \gamma_5 \right] \psi_f H_i \;, \label{eq:yuklag}
\eeq
where $\psi_f$ denote the fermion fields with mass $m_f$. The
coefficients of the CP-even and of the CP-odd part of the Yukawa
coupling, respectively, $c^e(H_i ff)$ and $c^o (H_i ff)$, have been
given in \cite{Fontes:2014xva} and
we repeat them here for convenience in table~\ref{tab:yukcoup}.
\begin{table}
\begin{center}
\begin{tabular}{rccc} \toprule
& $u$-type & $d$-type & leptons \\ \midrule
type I & $\frac{R_{i2}}{s_\beta} - i \frac{R_{i3}}{t_\beta} \gamma_5$
& $\frac{R_{i2}}{s_\beta} + i \frac{R_{i3}}{t_\beta} \gamma_5$ &
$\frac{R_{i2}}{s_\beta} + i \frac{R_{i3}}{t_\beta} \gamma_5$ \\
type II & $\frac{R_{i2}}{s_\beta} - i \frac{R_{i3}}{t_\beta} \gamma_5$
& $\frac{R_{i1}}{c_\beta} - i t_\beta R_{i3} \gamma_5$ &
$\frac{R_{i1}}{c_\beta} - i t_\beta R_{i3} \gamma_5$ \\
lepton-specific & $\frac{R_{i2}}{s_\beta} - i \frac{R_{i3}}{t_\beta} \gamma_5$
& $\frac{R_{i2}}{s_\beta} + i \frac{R_{i3}}{t_\beta} \gamma_5$ &
$\frac{R_{i1}}{c_\beta} - i t_\beta R_{i3} \gamma_5$ \\
flipped & $\frac{R_{i2}}{s_\beta} - i \frac{R_{i3}}{t_\beta} \gamma_5$
& $\frac{R_{i1}}{c_\beta} - i t_\beta R_{i3} \gamma_5$ &
$\frac{R_{i2}}{s_\beta} + i \frac{R_{i3}}{t_\beta} \gamma_5$ \\ \bottomrule
\end{tabular}
\caption{Coupling coefficients of the Yukawa couplings of the Higgs
  bosons $H_i$ in the C2HDM. The expressions correspond to
  $[c^e(H_i ff) +i c^o (H_i ff) \gamma_5]$ from
  Eq.~(\ref{eq:yuklag}). \label{tab:yukcoup}}
\end{center}
\end{table}
Further Higgs couplings of the C2HDM can be found in
\cite{Fontes:2014xva}. We implemented the C2HDM in the Fortran code
{\tt HDECAY}. This version of the program, which provides the Higgs
decay widths and
branching ratios of the C2HDM including the state-of-the-art higher
order QCD corrections and off-shell decays, will be released in a
future publication.

\subsection{The N2HDM\label{sec:n2hdm}}
In a recent publication \cite{Muhlleitner:2016mzt} we have studied the
phenomenology
of the N2HDM including the theoretical and experimental
constraints. We presented there for the first time a systematic
analysis of the global minimum of the N2HDM. For details on this
analysis and the tests of tree-level perturbativity and
vacuum stability we refer to \cite{Muhlleitner:2016mzt}. We restrict ourselves here
to briefly introducing the model. \s

The N2HDM is obtained from the CP-conserving 2HDM with a softly broken
$\mathbb{Z}_2$ symmetry upon extension by a real singlet field
$\Phi_S$ with a discrete symmetry, $\Phi_S \to - \Phi_S$. The N2HDM
potential is given by
\beq
V &=& m_{11}^2 |\Phi_1|^2 + m_{22}^2 |\Phi_2|^2 - m_{12}^2 (\Phi_1^\dagger
\Phi_2 + h.c.) + \frac{\lambda_1}{2} (\Phi_1^\dagger \Phi_1)^2 +
\frac{\lambda_2}{2} (\Phi_2^\dagger \Phi_2)^2 \nonumber \\
&& + \lambda_3
(\Phi_1^\dagger \Phi_1) (\Phi_2^\dagger \Phi_2) + \lambda_4
(\Phi_1^\dagger \Phi_2) (\Phi_2^\dagger \Phi_1) + \frac{\lambda_5}{2}
[(\Phi_1^\dagger \Phi_2)^2 + h.c.] \nonumber \\
&& + \frac{1}{2} u_S^2 \Phi_S^2 + \frac{\lambda_6}{8} \Phi_S^4 +
\frac{\lambda_7}{2} (\Phi_1^\dagger \Phi_1) \Phi_S^2 +
\frac{\lambda_8}{2} (\Phi_2^\dagger \Phi_2) \Phi_S^2 \;.
\label{eq:n2hdmpot}
\eeq
The first two lines contain the 2HDM part and the last line the
contributions of the singlet field $\Phi_S$. Working in the
CP-conserving 2HDM, all parameters in (\ref{eq:n2hdmpot}) are real.
Extensions by a singlet field that does not acquire a VEV provide a
viable DM candidate
\cite{He:2008qm,Grzadkowski:2009iz,Logan:2010nw,Boucenna:2011hy,He:2011gc,Bai:2012nv,He:2013suk,Cai:2013zga,Wang:2014elb,Drozd:2014yla,Campbell:2015fra,vonBuddenbrock:2016rmr}. We
do not consider this option here. The
doublet and singlet fields after EWSB can be parametrized as
\beq
\Phi_1 = \left( \begin{array}{c} \phi_1^+ \\ \frac{1}{\sqrt{2}} (v_1 +
    \rho_1 + i \eta_1) \end{array} \right) \;, \quad
\Phi_2 = \left( \begin{array}{c} \phi_2^+ \\ \frac{1}{\sqrt{2}} (v_2 +
    \rho_2 + i \eta_2) \end{array} \right) \;, \quad
\Phi_S = v_S + \rho_S \;,
\eeq
where $v_{1,2}$ denote the VEVs of the doublets $\Phi_{1,2}$ and $v_S$
the singlet VEV. The minimum conditions of the potential lead to the
three conditions
\beq
\frac{v_2}{v_1} m_{12}^2 - m_{11}^2 &=& \frac{1}{2} (v_1^2 \lambda_1 +
v_2^2 \lambda_{345} + v_S^2 \lambda_7) \label{eq:n2hdmmin1} \\
\frac{v_1}{v_2} m_{12}^2 - m_{22}^2 &=& \frac{1}{2} (v_1^2 \lambda_{345} +
v_2^2 \lambda_2 + v_S^2 \lambda_8) \label{eq:n2hdmmin2} \\
- m_S^2 &=& \frac{1}{2} (v_1^2 \lambda_7 + v_2^2 \lambda_8 + v_S^2
\lambda_6) \;, \label{eq:n2hdmmin3}
\eeq
with
\beq
\lambda_{345} \equiv \lambda_3 + \lambda_4 + \lambda_5 \;.
\label{eq:l345}
\eeq
As usual the mass matrices in the gauge basis are obtained from the
second derivatives of the Higgs potential in the electroweak minimum with
respect to the fields in the gauge basis. As we do not allow for a
complex singlet VEV, the particle content of the charged and
pseudoscalar sectors do not change when compared to the real 2HDM, and
their mass matrices can be diagonalized through
\beq
R_\beta = \left( \begin{array}{cc} c_\beta & s_\beta \\ - s_\beta &
    c_\beta \end{array} \right) \;,
\eeq
with $t_\beta$ defined as in the C2HDM through $t_\beta =
v_2/v_1$. In the mass basis we are then left with the
charged and neutral would-be Goldstone
bosons $G^\pm$ and $G^0$ as well as the charged Higgs mass
eigenstates $H^\pm$ and the pseudoscalar mass eigenstate $A$. \s

The additional real singlet field induces  a $3\times 3$ mass matrix in the CP-even neutral sector, which in
the basis $(\rho_1, \rho_2, \rho_3\equiv \rho_S)$ can be cast into the form
\beq
M_{\text{scalar}}^2 = \left( \begin{array}{ccc} \lambda_1 c_\beta^2 v^2 + t_\beta
    m_{12}^2 & \lambda_{345} c_\beta s_\beta v^2 - m_{12}^2 &
    \lambda_7 c_\beta v v_S \\ \lambda_{345} c_\beta s_\beta v^2 &
    \lambda_2 s_\beta^2 v^2 + m_{12}^2/t_\beta & \lambda_8 s_\beta v
    v_S \\ \lambda_7 c_\beta v v_S & \lambda_8 s_\beta v v_S &
    \lambda_6 v_S^2 \end{array} \right) \;,
\eeq
where we have used Eqs.~(\ref{eq:n2hdmmin1})-(\ref{eq:n2hdmmin3}),  to
replace the mass parameters $m_{11}^2$, $m_{22}^2$ and $m_S^2$ by $v=
\sqrt{v_1^2+ v_2^2}$, $t_\beta$ and $v_S$. We parametrize the
orthogonal matrix $R$ that diagonalizes the mass matrix again as
in Eq.~(\ref{eq:singletmatrix}) in terms of the mixing angles
$\alpha_i$ with the same ranges as before, see \eqref{eg:alpharanges}.
The physical mass eigenstates $H_1$ to $H_3$ are related to the
interaction states $(\rho_1, \rho_2, \rho_3)$ through
\beq
\left( \begin{array}{c} H_1 \\ H_2 \\ H_3 \end{array} \right) = R
\left( \begin{array}{c} \rho_1 \\ \rho_2 \\ \rho_3 \end{array} \right) \; .
\label{eq:rotn2hdm}
\eeq
The diagonalized mass matrix $M_{\text{scalar}}^2$ is obtained as
\beq
R M_{\text{scalar}}^2 R^T = \mbox{diag}(m_{H_1}^2,m_{H_2}^2,m_{H_3}^2) \;,
\eeq
with the mass eigenstates ordered by ascending mass as
\beq
m_{H_1} < m_{H_2} < m_{H_3} \;.
\eeq

There are altogether 12 independent real parameters describing the
N2HDM, among which we choose as many parameters with physical
meaning as possible. We use the minimisation
conditions to replace $m_{11}^2$, $m_{22}^2$ and $m_S^2$ by the SM VEV,
$t_\beta$ and $v_S$. The quartic couplings are traded for the
physical masses and the mixing angles. Together with
the soft $\mathbb{Z}_2$ breaking parameter, our physical parameter
set reads
\beq
\alpha_1 \; , \quad \alpha_2 \; , \quad \alpha_3 \; , \quad t_\beta \;
, \quad v \; ,
\quad v_s \; , \quad m_{H_{1,2,3}} \;, \quad m_A \;, \quad m_{H^\pm}
\;, \quad m_{12}^2 \;.
\eeq
The expressions of the quartic couplings in terms of
the physical parameter set can be found in appendix~A.1 of
\cite{Muhlleitner:2016mzt}.\s

The singlet field $\rho_S$ does not couple directly to the SM
particles so that the only change in the tree-level Higgs boson
couplings with respect to the CP-conserving 2HDM is due to the mixing
of the three neutral fields $\rho_i$. Therefore, couplings that do not
involve the CP-even neutral Higgs bosons remain unchanged compared to
the 2HDM. They have been given {\it e.g.}~in \cite{Branco:2011iw}.
The problem of possible non-zero FCNC is solved by extending the
$\mathbb{Z}_2$ symmetry to the Yukawa sector, so that the same four
types of doublet couplings to the fermions are obtained as in the
2HDM. For the specific form of all relevant coupling factors we refer
to \cite{Muhlleitner:2016mzt}.

\subsection{The NMSSM \label{sec:nmssm}}
Supersymmetry requires the introduction of at least two Higgs
doublets. In the NMSSM a complex superfield $\hat{S}$ is added to this
minimal supersymmetric field content with the doublet superfields
$\hat{H}_u$ and $\hat{H}_d$. This allows for a dynamic solution of the
$\mu$ problem in the MSSM  when the singlet field acquires a
non-vanishing VEV. After EWSB the NMSSM Higgs spectrum comprises seven
physical Higgs states. In the CP-conserving case, investigated in this work, these
are three neutral CP-even, two neutral CP-odd and two charged Higgs
bosons. The NMSSM Higgs potential is obtained from the superpotential,
the soft SUSY breaking Lagrangian and the $D$-term contributions. In
terms of the hatted superfields the scale-invariant NMSSM
superpotential is
\beq
{\cal W} = \lambda \widehat{S} \widehat{H}_u \widehat{H}_d +
\frac{\kappa}{3} \, \widehat{S}^3 + h_t
\widehat{Q}_3\widehat{H}_u\widehat{t}_R^c - h_b \widehat{Q}_3
\widehat{H}_d\widehat{b}_R^c  - h_\tau \widehat{L}_3 \widehat{H}_d
\widehat{\tau}_R^c \; .
\label{eq:superpotential}
\eeq
We have included only the third generation fermion
superfields here as an example. These are the left-handed doublet quark
($\widehat{Q}_3$) and lepton ($\widehat{L}_e$) superfields as well as the
right-handed singlet quark ($\widehat{t}^c_R,\widehat{b}^c_R$) and lepton
($\widehat{\tau}^c_R$) superfields. The first term in Eq.~(\ref{eq:superpotential})
replaces the $\mu$-term $\mu \hat{H}_d \hat{H}_u$
of the MSSM superpotential, the term cubic in the singlet superfield
breaks the Peccei-Quinn symmetry thus preventing the appearance of a
massless axion and the last three terms describe the Yukawa
interactions. The soft SUSY breaking Lagrangian contains
contributions from the mass terms for the Higgs and the sfermion
fields, that are built from the complex scalar components of the superfields, i.e.
\beq
\label{eq:Lagmass}
 -{\cal L}_{\mathrm{mass}} &=&
 m_{H_u}^2 | H_u |^2 + m_{H_d}^2 | H_d|^2 + m_{S}^2| S |^2 \nonumber \\
  &+& m_{{\tilde Q}_3}^2|{\tilde Q}_3^2| + m_{\tilde t_R}^2 |{\tilde t}_R^2|
 +  m_{\tilde b_R}^2|{\tilde b}_R^2| +m_{{\tilde L}_3}^2|{\tilde L}_3^2| +
 m_{\tilde  \tau_R}^2|{\tilde \tau}_R^2|\; .
\eeq
The soft SUSY breaking part with the trilinear soft SUSY breaking
interactions between the sfermions and the Higgs fields is given by
\beq
\label{eq:Trilmass}
-{\cal L}_{\mathrm{tril}}=  \lambda A_\lambda H_u H_d S + \frac{1}{3}
\kappa  A_\kappa S^3 + h_t A_t \tilde Q_3 H_u \tilde t_R^c - h_b A_b
\tilde Q_3 H_d \tilde b_R^c - h_\tau A_\tau \tilde L_3 H_d \tilde \tau_R^c
+ \mathrm{h.c.} \;
\eeq
with the $A$'s denoting the soft SUSY breaking trilinear couplings.
Soft SUSY breaking due to the gaugino mass parameters $M_{1,2,3}$ of
the bino ($\tilde{B}$), winos ($\tilde{W}$) and gluinos ($\tilde{G}$), respectively,
is described by
\beq
-{\cal L}_\mathrm{gauginos}= \frac{1}{2} \bigg[ M_1 \tilde{B}
\tilde{B}+M_2 \sum_{a=1}^3 \tilde{W}^a \tilde{W}_a +
M_3 \sum_{a=1}^8 \tilde{G}^a \tilde{G}_a  \ + \ {\rm h.c.}
\bigg] \;.
\eeq
We will allow for non-universal soft terms at the GUT scale. \s

After EWSB we expand the tree-level scalar potential around the
non-vanishing VEVs of the Higgs doublet and singlet fields,
\beq
H_d = \left( \begin{array}{c} (v_d + h_d + i a_d)/\sqrt{2} \\
   h_d^- \end{array} \right) \,, \;
H_u = \left( \begin{array}{c} h_u^+ \\ (v_u + h_u + i a_u)/\sqrt{2}
 \end{array} \right) \,, \;
S= \frac{v_s+h_s+ia_s}{\sqrt{2}}\;.
\eeq
We obtain the Higgs mass matrices for the three scalars ($h_d, h_u, h_s$), the three pseudoscalars ($a_d,a_u,a_s$) and the charged Higgs states
($h_u^\pm,h_d^\mp$) from the second derivative of the scalar potential. We choose the VEVs $v_u, v_d$ and $v_s$ to be real and
positive. The CP-even mass eigenstates $H_i$ ($i=1,2,3$) are obtained
through a rotation with the orthogonal matrix ${\cal R}^S$
\beq
(H_1, H_2, H_3)^T = {\cal R}^S (h_d,h_u,h_s)^T \;,
\label{eq:scalarrotation}
\eeq
which diagonalizes the $3\times 3$ mass matrix squared, $M^2_S$, of
the CP-even fields. The mass eigenstates are ordered by ascending mass,
$M_{H_1} \le M_{H_2} \le M_{H_3}$. The CP-odd mass eigenstates $A_1$
and $A_2$ are obtained by performing first a rotation ${\cal R}^G$ to
separate the massless Goldstone boson and then a rotation ${\cal R}^P$
into the mass eigenstates,
\beq
(A_1,A_2,G)^T = {\cal R}^P {\cal R}^G (a_d,a_u,a_s)^T \;,
\label{eq:pseudorot}
\eeq
which are also ordered by ascending mass, $M_{A_1} \le M_{A_2}$. \s

We use the three minimisation conditions of the scalar potential to
express the soft SUSY breaking masses squared for $H_u$, $H_d$ and $S$
in ${\cal L}_{\text{mass}}$ in terms of the remaining parameters of the
tree-level scalar potential. The tree-level NMSSM Higgs sector can
hence be parametrized in terms of the six parameters
\beq
\lambda\ , \ \kappa\ , \ A_{\lambda} \ , \ A_{\kappa}, \
\tan \beta =v_u/ v_d \quad \mathrm{and}
\quad \mu_\mathrm{eff} = \lambda v_s/\sqrt{2}\; .
\eeq
The sign conventions are chosen such that $\lambda$ and $\tan\beta$
are positive, whereas $\kappa, A_\lambda, A_\kappa$ and
$\mu_{\text{eff}}$ can have both signs. Note that the Higgs boson
masses, in contrast to the non-SUSY Higgs sector extensions discussed
in this work, are not input parameters but have to be calculated
including higher order corrections. The latter is crucial in order to
obtain a realistic mass prediction for the SM-like Higgs mass, which
is measured to be 125~GeV. Through these corrections also the
soft SUSY breaking mass terms for the scalars and the gauginos as well
as the trilinear soft SUSY breaking couplings enter the Higgs
sector. Another difference to the other BSM Higgs
sectors is that the parameters have to respect SUSY
relations with significant phenomenological consequences.

\section{Parameter Scans \label{sec:scans}}
In order to perform phenomenological analyses with the presented
models we need viable parameter points, {\it
  i.e.}~points in accordance with theoretical and experimental
constraints. To obtain these points we perform extensive scans in the
parameter space of each model and check for compatibility with the
constraints. In case of the CxSM, C2DHM and N2HDM this is done by
using the program {\tt ScannerS}. The phenomenology of the C2HDM and
N2HDM also depends on the treatment of the Yukawa sector. We will
focus our discussion on the examples of type I and type II models.
In the following we denote the
discovered SM-like Higgs boson by $h_{125}$ with a mass of \cite{Aad:2015zhl}
\beq
m_{h_{125}} = 125.09 \; \mbox{GeV}\;.
\eeq
In all models we exclude parameter configurations where
the Higgs signal is built up by two resonances. To this end we demand
the mass window $m_{h_{125}} \pm 5$~GeV to be free of any Higgs bosons except for
$h_{125}$.  We fix the doublet VEV $v$ to the SM
value. Furthermore, we do not include electroweak
corrections in the parameter scans nor in the analysis, as they are
not (entirely) available for all models and cannot be taken over from
the SM.

\subsection{The CxSM Parameter Scan  \label{sec:cxsmscan}}
In the CxSM we re-used the sample generated for
\cite{Costa:2015llh}. We briefly repeat the constraints that
have been applied and refer to \cite{Costa:2015llh} for details. The
applied theoretical constraints are the requirement on the potential to
be bounded from below, that the chosen vacuum is a global minimum and
that perturbative unitarity holds. The compatibility with the
electroweak precision data has been ensured by applying a 95\% C.L.~exclusion
limit from the electroweak precision observables $S$, $T$ and $U$
\cite{Peskin:1991sw,Maksymyk:1993zm}, see
\cite{Costa:2014qga} for further information.
The 95\% C.L.~exclusion limits from the LHC Higgs data have been
applied by using {\tt HiggsBounds} \cite{Bechtle:2013wla}. We then
keep only those parameter points where the $h_{125}$ is in accordance with
the Higgs data by requiring that the global signal strength $\mu$ is
within $2\sigma$ of the experimental fit value
\cite{Khachatryan:2016vau}.\footnote{In adopting this procedure we are
  allowing a larger number of points 
in our sample than the ones that would be obtained if we considered the
six-dimensional ellipsoid. We are in fact considering the points that
are inside the bounding box of this ellipsoid.
Moreover, we also overestimate the allowed range by considering $2 \times 1\sigma$
instead of $2\sigma$. One should note that this is a preliminary study comparing the
phenomenology of several models and that the procedure is the same for
all models.} 
With the mixing matrix $R$ defined in Eq.~(\ref{eq:rotsinglet}) we calculate
$\mu$, at leading order in the electroweak parameters, as
\beq
\mu = (R_{h_{125} \, 1})^2 \times \Sigma_{X_{\text{SM}}}
BR(h_{125} \to X_{\text{SM}}) \;,
\eeq
where $X_{\text{SM}}$ denotes a SM particle pair final state and $i$
refers to that of the $H_i$ in Eq.~(\ref{eq:rotsinglet}) that is
identified with the $h_{125}$. The
branching ratios have been obtained with the Fortran code {\tt
  sHDECAY} \cite{Costa:2015llh}.
We do not include the effects of chain production \cite{Costa:2015llh} here nor
in any of the other models. \s

The sample was generated with the input parameters given in
Eq.~(\ref{eq:cxsminput}). One of the Higgs bosons is identified with
$h_{125}$ and the remaining ones are restricted to the mass range
\beq
30 \; \mbox{GeV } \le m_{H_i} < 1000 \;\mbox{GeV} ,\;H_i \ne h_{125} \;.
\eeq
The VEVs $v_A$ and $v_S$ are varied in the range
\beq
1 \mbox{ GeV } \le v_A, v_S < 1.5 \mbox{ TeV} \;.
\eeq
The mixing angles $\alpha_{1,2,3}$ are chosen in
\beq
-\frac{\pi}{2} \le \alpha_{1,2,3} < \frac{\pi}{2} \;.
\eeq
All input parameters were randomly generated (uniformly) in the ranges specified above and we obtained $\sim 4\times 10^6$ valid points.

\subsection{The C2HDM Parameter Scan  \label{sec:2hdmscan}}
We have implemented the C2DHM as a {\tt ScannerS} model class. This
allowed us to perform a full parameter space scan that simultaneously
applies the constraints described here: We require the
potential to be bounded from below and we use the  tree-level discriminant from
\cite{Ivanov:2015nea} to enforce that the vacuum configuration is at a global minimum to disallow vacuum decay. Furthermore, we check that tree-level
perturbative unitarity holds. We apply the flavour constraints on $R_b$
\cite{Haber:1999zh,Deschamps:2009rh} and
$B \to X_s \gamma$ \cite{Deschamps:2009rh,Mahmoudi:2009zx,Hermann:2012fc,Misiak:2015xwa,Misiak:2017bgg}, which can be generalized from the
CP-conserving 2HDM to the C2HDM as they only depend on the charged Higgs
boson. These constraints are checked as $2\sigma$ exclusion bounds on
the $m_{H^\pm}-t_\beta$ plane. Note that the latest calculation
of Ref.~\cite{Misiak:2017bgg} enforces
\beq
m_{H^\pm} > 580 \mbox{ GeV}
\eeq
in the type II and flipped 2HDM. In the type I model this
bound is much weaker and depends more strongly on $\tan\beta$.
We verify agreement with the
electroweak precision measurements by using the oblique parameters
$S$, $T$ and $U$. The formulae for their computation in the general
2HDM have been given in \cite{Branco:2011iw}. For the computed $S$, $T$ and $U$
values we demand $2\sigma$ compatibility with the SM fit \cite{Baak:2014ora}. The
full correlation among the three parameters is taken into
account. Again, compatibility with the Higgs data is checked using
{\tt HiggsBounds}\footnote{
A recent ATLAS analysis~\cite{ATLAS:2016pyq} 
considered a pseudoscalar of mass $500\,\text{GeV}$ 
decaying into a $t\bar t$-pair. Assuming a type II 2HDM, 
they obtained a constraint of $\tan\beta>0.85$ 
for a pseudoscalar of this mass. Although relevant, this
constraint can only be applied in the immediate vicinity of 
a pseudoscalar mass of 500 GeV and therefore we did not include 
it in our analysis.} and the individual signal strengths fit
\cite{Khachatryan:2016vau} for the $h_{125}$.
The necessary decay widths and
branching ratios are obtained from a private implementation of the
C2HDM into {\tt HDECAY v6.51}, which will be released in a future
publication. This includes state-of-the-art QCD corrections and
off-shell decays. Additionally we need the Higgs boson production cross
sections normalized to the SM. The gluon fusion ($ggF$) and $b$-quark
fusion ($bbF$) production cross sections at next-to-next-to-leading
order (NNLO) QCD are obtained from {\tt SusHi
  v1.6.0} \cite{Harlander:2012pb,Harlander:2016hcx} which is interfaced with {\tt
  ScannerS}. The cross section contributions from the CP-even and the
CP-odd Yukawa couplings are calculated separately and then added
incoherently. Hence, the fermion initiated cross section normalized to the SM
is given by
\beq
\mu_F = \frac{\sigma^{\text{even}}_{\text{C2HDM}} (ggF)
  +\sigma^{\text{even}}_{\text{C2HDM}} (bbF)
  +\sigma^{\text{odd}}_{\text{C2HDM}} (ggF) +
  \sigma^{\text{odd}}_{\text{C2HDM}}
  (bbF)}{\sigma^{\text{even}}_{\text{SM}} (ggF)} \;.
\eeq
In the denominator we neglected the $bbF$ cross section which is very small compared to gluon fusion production in
the SM. The QCD
corrections to massive gauge boson-mediated production cross sections
cancel upon normalization to the SM. Thus, vector boson fusion ($VBF$) and
associated production with vector bosons ($VH$) yield the normalized production strength
\beq
\mu_V = \frac{\sigma^{\text{even}}_{\text{C2HDM}}
  (VBF)}{\sigma^{\text{even}}_{\text{SM}} (VBF)} =
\frac{\sigma^{\text{even}}_{\text{C2HDM}}
  (VH)}{\sigma^{\text{even}}_{\text{SM}} (VH)} = c^2 (H_i VV) \;,
\eeq
with the effective coupling defined in
Eq.~(\ref{eq:gaugecoupdef}). There are, obviously, no CP-odd
contributions to these channels (at tree-level). {\tt HiggsBounds} also requires the
cross sections for associated production with top or bottom
quarks. Due to the different QCD corrections of the CP-even and CP-odd
contributions to these processes \cite{Hafliger:2005aj}, the QCD corrections in their
incoherent addition do not cancel when normalized to the SM.
Therefore, we use these
cross section ratios only at leading order. The ratios are given
by
\beq
\mu_{\text{assoc}} = \frac{\sigma_{\text{C2HDM}}
  (ffH_i)}{\sigma_{\text{SM}} (ffH)} = c^e (H_i ff)^2 + c^o (H_i ff)^2 \;,
\eeq
with the coupling coefficients defined in Eq.~(\ref{eq:yuklag}). This
information is passed to {\tt HiggsBounds} via the {\tt ScannerS}
interface and {\tt HiggsBounds v4.3.1} is used to check agreement with
all $2\sigma$ exclusion limits from LEP, Tevatron and LHC Higgs
searches. The properties of the $h_{125}$ are checked against the fitted
values of
\beq
\frac{\mu_F}{\mu_V} \;, \quad \mu_{\gamma\gamma} \;, \quad
\mu_{ZZ} \;, \quad \mu_{WW} \;, \quad  \mu_{\tau\tau} \;, \quad
\mu_{bb} \;,
\eeq
given in \cite{Khachatryan:2016vau}, with $\mu_{xx}$ defined as
\beq
\mu_{xx} = \mu_F \, \frac{\mbox{BR}_{\text{C2HDM}} (H_i \to
  xx)}{\mbox{BR}_{\text{SM}} (H_i \to xx)}
\eeq
for $H_i \equiv h_{125}$.
We require agreement with the fit results of \cite{Khachatryan:2016vau} within
the $2\times 1\sigma$ level. All our models preserve custodial symmetry so that
\beq
\mu_{ZZ} = \mu_{WW} \equiv \mu_{VV} \;.
\eeq
Therefore, we combine the lower $2 \times 1\sigma$ bound from
$\mu_{ZZ}$ with the upper bound on $\mu_{WW}$ \cite{Khachatryan:2016vau} and use
\beq
0.79 < \mu_{VV} < 1.48 \;.
\eeq
Strong constraints on CP violation in the Higgs sector arise
from electric dipole moment (EDM) measurements, among which the
one of the electron imposes the strongest constraints \cite{Inoue:2014nva}, with the
experimental limit given by the ACME collaboration
\cite{Baron:2013eja}. We have implemented the calculation of
\cite{Abe:2013qla} and applied the constraints from the electron EDM
in a full scan of the C2HDM parameter space.
We require our results to be compatible with the values given in
\cite{Baron:2013eja} at 90\% C.L. \s

For the scan with the input parameters from
Eq.~(\ref{eq:2hdminputset}) we choose $t_\beta$ in the range
\beq
0.25 \le t_\beta \le 35 \;. \label{eq:tbscanc2hdm}
\eeq
As the lower bound on $t_\beta$ from the $R_b$ measurement is stronger
than the lower bound in Eq.~(\ref{eq:tbscanc2hdm}), the latter has no
influence on the physical parameter points. After transforming the
mixing matrix generated by {\tt ScannerS} to the parametrization of
Eq.~(\ref{eq:singletmatrix}) we allow the mixing angles to vary in
\beq
- \frac{\pi}{2} \le \alpha_{1,2,3} < \frac{\pi}{2} \;.
\eeq
The value of $\mbox{Re} (m_{12}^2)$ is chosen in
\beq
0 \mbox{ GeV}^2 \le \mbox{Re}(m_{12}^2)  < 500 000 \mbox{ GeV}^2 \;.
\eeq
There are also physical parameter points with $\mbox{Re}(m_{12}^2) <0$
but they are extremely rare, and we neglect them in our study.
We identify one of the neutral Higgs bosons $H_i$ with $h_{125}$. In
type II, the charged Higgs mass is chosen in the range
\beq
580 \mbox{ GeV } \le m_{H^\pm} < 1 \mbox{ TeV } \;,
\eeq
and in type I we choose
\beq
80 \mbox{ GeV } \le m_{H^\pm} < 1 \mbox{ TeV } \;.
\eeq
The electroweak precision constraints combined with perturbative unitarity constraints force the mass of at
least one of the neutral Higgs bosons to be close to $m_{H^\pm}$.
Therefore, we increase the efficiency of the parameter scan
by generating a second neutral Higgs mass $m_{H_i \ne h_{125}}$ in the
interval
\beq
500 \mbox{ GeV}\leq m_{H_i}<1 \mbox{ TeV}
\eeq
in the type II and
\beq
30 \mbox{ GeV}\leq m_{H_i}<1 \mbox{ TeV}
\eeq
in the type I.
The third neutral Higgs
boson $m_{H_j \ne H_i, h_{125}}$ is not an independent parameter and is
calculated by {\tt ScannerS}. We require the masses of both Higgs
bosons $H_i, H_j \ne h_{125}$ to lie in the interval
\beq
30 \mbox{ GeV } \le m_{H_i}, m_{H_j} < 1 \mbox{ TeV } .
\eeq
We have generated samples of $\sim10^5$ valid points within these
bounds for type I and for type II. Since we found the CP-conserving
limit not to be well-captured by this scan we added another $\sim10^5$
CP-conserving points to each of these samples ($\sim8\times 10^4$
points where $h=h_{125}$ and $\sim2\times10^4$ points where
$H=h_{125}$). These points were generated in the same ranges and with
the same constraints applied.\footnote{Except for the EDM constraint which is trivially satisfied if CP is conserved.}

\subsection{The N2HDM Parameter Scan  \label{sec:n2hdmscan}}
We check for the theoretical constraints, namely that the potential is
bounded from below, that the chosen vacuum is the global minimum and
that perturbative unitarity holds, as described in detail in
\cite{Muhlleitner:2016mzt}. \s

Most of the experimental constraints applied on the C2HDM described in
section~\ref{sec:2hdmscan} are also valid for the N2HDM. Since the
constraints on $R_b$ \cite{Haber:1999zh,Deschamps:2009rh} and $B \to X_s \gamma$
\cite{Deschamps:2009rh,Mahmoudi:2009zx,Hermann:2012fc,Misiak:2015xwa,Misiak:2017bgg}
are only sensitive to the charged Higgs boson
the 2HDM calculation and the resulting $2\sigma$ limits in the
$m_{H^\pm} - t_\beta$ plane can also be used in the N2HDM. For the
oblique parameters $S$, $T$ and $U$, calculated with the general
formulae in \cite{Grimus:2007if,Grimus:2008nb},
$2\sigma$ compatibility
with the SM fit~\cite{Baak:2014ora} including the full correlations is
demanded.
The check of compatibility with the Higgs data proceeds analogously
to the one described for the C2HDM modulo the different Higgs spectrum
to be investigated and the replacement of the production cross
sections in the signal rates with the corresponding ones for the production of
either a purely CP-even or a purely CP-odd N2HDM Higgs boson. \s

For the scan we choose the following parameter ranges
\beq
\begin{array}{ll}
-\frac{\pi}{2} \le \alpha_{1,2,3} < \frac{\pi}{2}\;, &
0.25 \le t_\beta \le 35\;, \\[0.2cm]
0 \mbox{ GeV}^2 \le \mbox{Re}(m_{12}^2) < 500000 \mbox{ GeV}^2\;, &
1 \mbox{ GeV} \le v_S \le 1.5 \mbox{ TeV}\;, \\[0.2cm]
30 \mbox{ GeV} \le m_{H_i \ne m_{h_{125}}}, m_A \le 1 \mbox{ TeV}\;, \\[0.2cm]
80 \mbox{ GeV} \le m_{H^\pm} < 1 \mbox{ TeV (type I)}\;, &
580 \mbox{ GeV} \le m_{H^\pm} < 1 \mbox{ TeV (type II)} \;.
\end{array}
\eeq
Within these ranges we generated samples of $\sim 5\times 10^5$ valid points for each type.

\subsection{The NMSSM Parameter Scan  \label{sec:nmssmscan}}
For the NMSSM parameter scan we follow the procedure described in
\cite{Costa:2015llh,King:2014xwa} and briefly summarise the main
features. The {\tt NMSSMTools} package
\cite{Ellwanger:2004xm,Ellwanger:2005dv,Ellwanger:2006rn,Das:2011dg,Muhlleitner:2003vg,Belanger:2005kh}
is used to compute the
spectrum of the Higgs and SUSY particles including higher order
corrections and to check for vacuum stability, the constraints from
low-energy observables and to compute the input required by {\tt
  HiggsBounds} to verify compatibility with the exclusion bounds from
Higgs searches.
The Higgs branching ratios of {\tt NMSSMTools} are cross-checked against {\tt
  NMSSMCALC} \cite{Baglio:2013iia}. The relic density is obtained via an interface
with {\tt micrOMEGAS} \cite{Belanger:2005kh} and required not to exceed the value
measured by the PLANCK collaboration \cite{Ade:2013zuv}.
We also obtained the spin-independent nucleon-dark matter direct
detection cross section using micrOMEGAS and required that it does not
violate the upper bound from the LUX
experiment~\cite{Akerib:2016vxi}. Only those parameter
points are retained that feature a neutral CP-even Higgs boson with
mass between 124 and 126 GeV. For this Higgs boson agreement with the
signal strength fit of \cite{Khachatryan:2016vau} is required at the
$2 \times 1\sigma$ level. For the gluon fusion cross section the ratio
between the NMSSM Higgs decay
width into gluons and the corresponding SM decay width at the same
mass value is multiplied with the SM gluon fusion cross section. The
branching ratios are taken from {\tt NMSSMTools} at NLO QCD, whereas the SM
cross section was calculated at NNLO QCD with {\tt HIGLU} \cite{Spira:1995mt}.
The cross section for $b\bar{b}$ annihilation is obtained from the
multiplication of the SM cross section with the effective squared
$b\bar{b}$ coupling of {\tt NMSSMTools}. For the SM
cross section values we use the ones from \cite{Ferreira:2014dya}
produced with the code {\tt  SusHi} \cite{Harlander:2012pb,Harlander:2016hcx}.
Furthermore, the obtained parameter points are checked for
compatibility with the SUSY searches at
  LHC\footnote{We take the limits given by the ATLAS
      collaboration. Comparable results were obtained by the CMS
      collaboration.} and the lower
bound on the charged Higgs mass \cite{TheATLAScollaboration:2013wia,CMS:mxa}.
Since the SUSY limits are model-dependent, we
decided to take them into account by applying conservative lower mass
limits.
On the masses of the gluinos and squarks of the first two generations
we imposed a lower bound of 1850~GeV \cite{Aad:2015iea}. We required the masses
of the lightest stop and sbottom to be heavier than 800 GeV
\cite{Aaboud:2016lwz,Aaboud:2016nwl}.\footnote{The mass of the
  lightest stop could also be considerably lighter in case the mass
  difference between the stop and the lightest neutralino is small
  \cite{Muhlleitner:2011ww,Grober:2014aha,Grober:2015fia,Aad:2015pfx,Aaboud:2016tnv}. Since this limit is model-dependent, we do not
  further take into account this case here.} Based on \cite{Aad:2014vma} we chose a
lower charged slepton mass limit of 400~GeV, and we required the lightest
chargino mass to be above 300~GeV \cite{Aad:2015jqa}. We did not impose
an extra cut on the neutralino mass, which would also depend on the
mass of the lightest chargino. Instead, the neutralino mass is constrained
by DM observables. 
\s

The ranges applied in our parameter scan are summarised in
table~\ref{tab:nmssmscan}. In order to ensure perturbativity we apply
the rough constraint
\beq
\lambda^2 + \kappa^2 < 0.7^2 \;.
\eeq
The remaining mass parameters of the third generation sfermions not
listed in the table are chosen as
\beq
m_{\tilde{t}_R} = m_{\tilde{Q}_3} \;, \quad m_{\tilde{\tau}_R} =
m_{\tilde{L}_3} \quad \mbox{ and } \; m_{\tilde{b}_R} = 3 \mbox{ TeV} \;.
\eeq
The mass parameters of the first and second generation sfermions are
set to 3 TeV. For consistency with the parameter ranges of the other
models we kept only points with all Higgs masses between 30~GeV and
1~TeV.
\begin{table}
\begin{center}
\begin{tabular}{l|ccc|cccccccccccc} \toprule
& $t_\beta$ & $\lambda$ & $\kappa$ & $M_1$ & $M_2$ & $M_3$ & $A_t$ &
$A_b$ & $A_\tau$ & $m_{\tilde{Q}_3}$ & $m_{\tilde{L}_3}$ & $A_\lambda$
& $A_\kappa$ & $\mu_{\text{eff}}$ \\
& & & & \multicolumn{11}{c}{in TeV} \\ \midrule
min & 1 & 0 & -0.7 & 0.1 & 0.2 & 1.3 & -2 & -2 & -2 & 0.6 & 0.6 & -2 &
-2 & -1 \\
max & 30 & 0.7 & 0.7 & 1 & 1 & 3 & 2 & 2 & 2 & 3 & 3 & 2 & 2 & 1 \\ \bottomrule
\end{tabular}
\caption{Input parameters for the NMSSM scan. All parameters have been
varied independently between the given minimum and maximum values. \label{tab:nmssmscan}}
\end{center}
\end{table}

With these constraints we performed a uniform scan of the parameters
within the boxes of table~\ref{tab:nmssmscan}. To improve the
efficiency of the scan, in a first step we check if a Higgs boson with
a tree-level mass inside the window  $125 \pm 100$~GeV is
present. Otherwise we reject the point before running it through {\tt
  NMSSMTools}. In the second step, after {\tt NMSSMTools} returns the loop corrected spectrum, we enforce that a Higgs boson is present with a mass inside the window $125\pm 1$~GeV. We also did part of the scan without this constraint
applied to ensure that we do not exclude more extreme scenarios with
larger radiative corrections. With this approach we obtained $\sim
7000$ valid points. 
 
\section{Phenomenological Analysis \label{sec:pheno}}
We now turn to our phenomenological analysis in which we study the
properties of the various models with the aim to identify features
unique to a specific model that allow us to distinguish between the
models. In our analysis of the C2HDM and the N2HDM we adopt the
most commonly studied type I and II Yukawa sectors.

\subsection{The Higgs Mass Spectrum}
\begin{figure}[t!]
  \centering
  \includegraphics[width=\linewidth]{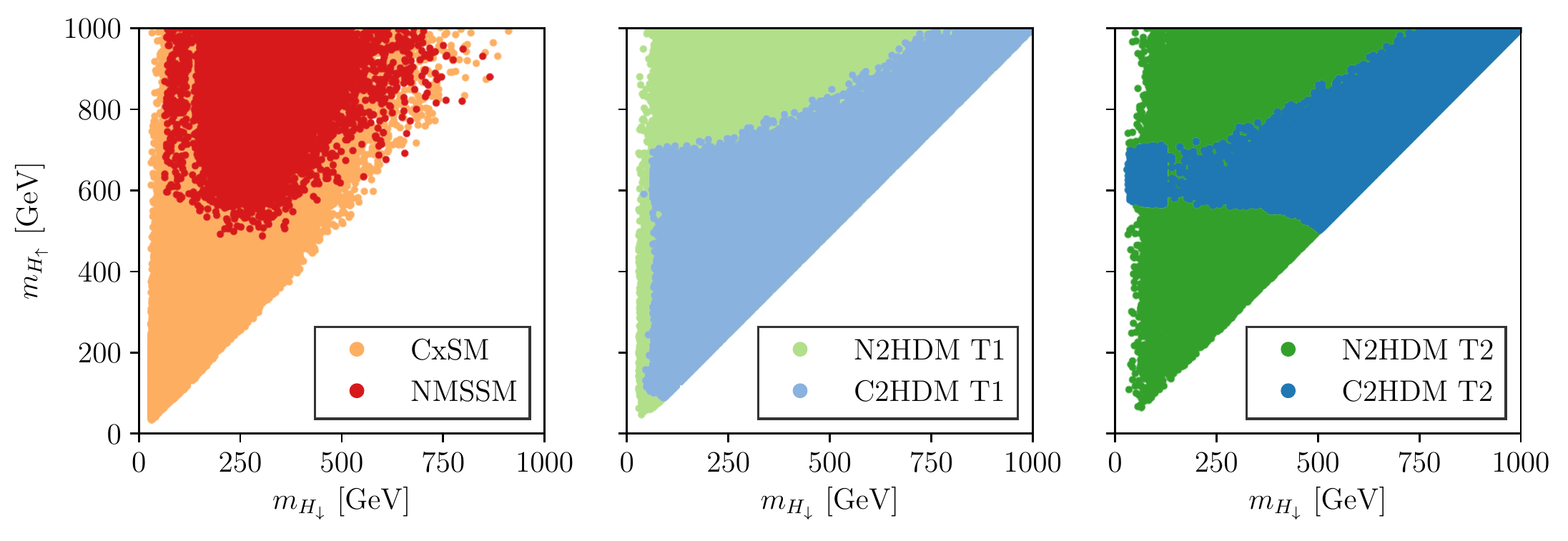}
  \caption{Masses of the two non-$h_{125}$ neutral scalars. Left:
    the CxSM (orange) and NMSSM (red); middle: the type I N2HDM (fair-green)
    and C2HDM (fair-blue); right: the type II N2HDM (dark-green) and
    C2HDM (dark-blue). For the CxSM, NMSSM and N2HDM
    the two masses in the axes are for CP-even Higgs bosons, whereas for the C2HDM they are for CP-mixed Higgs bosons .
    By definition $m_{H_\downarrow}\leq
    m_{H_\uparrow}$.} \label{fig:massdistr}
\end{figure}

We start the phenomenological comparison of our models by
investigating the Higgs mass spectrum. In
Fig.~\ref{fig:massdistr} we show for the CxSM, NMSSM
and N2HDM the mass distributions of the two neutral CP-even non-$h_{125}$ Higgs
bosons and for the C2HDM the ones of the two CP-mixed non-$h_{125}$
Higgs bosons. For the N2HDM and C2HDM we show the results both for
type I and type II. From now on we call the lighter of these $H_\downarrow$ and the
heavier one $H_\uparrow$. The N2HDM and NMSSM feature
additional CP-odd Higgs bosons. For the C2HDM, all our plots shown here and in the
following include the limit of the real 2HDM through a dedicated scan
in that model to improve the density. We have performed a lower
density localised scan of that region in the C2HDM to check that this
is consistent.
In all models we find points with $m_{H_\downarrow} < m_{h_{125}}$.
For the C2HDM type II, however, this is only the case in the limit of
the real 2HDM. Away from this limit the masses of $H_\downarrow$ and
$H_\uparrow$ turn out to be always heavier than about
500~GeV and to be close. We have verified that this results from a combination of the tree-level unitarity constraints with the electroweak precision data constraints (through the $S,T,U$ variables). To conclude this, first we performed several scans, one for each constraint with only that constraint applied, to check the individual effect of each constraint. Then we repeated the procedure for all possible pairings of constraints. The upper boundary of the C2HDM mass spectra observed in the middle and right panels is the same for both types and it matches the one for the real 2HDM. This boundary is due to tree-level unitarity constraints. 
In the N2HDM there is more freedom, with further quartic couplings involving the singlet, so the same boundary does not arise.
\s

In the N2HDM, the CxSM and C2HDM type I, we have points where
$m_{H^\uparrow} < m_{h_{125}}$ and hence the $h_{125}$ is the heaviest
of the CP-even (CP-mixed in the C2HDM) neutral Higgs
bosons. In our scan, we did not find such points for
  the NMSSM.\footnote{For a recent investigation of the NMSSM in view of the present Higgs data and
a discussion of the mass hierarchies, see~\cite{Domingo:2015eea,
  Carena:2015moc}.} The N2HDM and NMSSM 
feature additionally pseudoscalars that
can also be lighter than 125~GeV.
The N2HDM covers the largest mass region. With the largest number of parameters, not
restricted by additional supersymmetric symmetries, it is easiest in
this model to adjust it to be compatible with all the applied
constraints. Note, finally, that the gaps at 125~GeV are due to the
mass windows around $h_{125}$ in
order to avoid degenerate Higgs signals.

\subsection{Phenomenology of the Singlet or Pseudoscalar Admixture in $h_{125}$}
We investigate the phenomenology of the $h_{125}$ with respect to its
possible singlet or pseudoscalar admixture. In particular, we study to which extent
this influences the signal strengths of the $h_{125}$ and if this can
be used to distinguish between the models. Additionally, we compare
the CP-conserving singlet admixture with the CP-violating pseudoscalar
admixture. Since the measurement of CP violation is experimentally
very challenging\footnote{For recent experimental
    analyses, see \cite{Khachatryan:2016tnr,Aad:2016nal}.}, a $h_{125}$ of the
C2HDM could be misidentified as a CP-even Higgs boson in the present phase of the LHC. 
Moreover, since the Higgs couplings to gauge bosons have the same Lorentz structure
as the SM Higgs boson, a clear signal of CP violation would have to be seen either
via the couplings to fermions or via particular combinations of decays if other
Higgs were discovered \cite{Fontes:2015xva}.
A comparison of the singlet and pseudoscalar admixture is therefore appropriate. \s

In the CxSM, the singlet admixture to a Higgs boson $H_i$
is given by the sum of the real and complex singlet parts squared, {\it i.e.}
\beq
\Sigma_i^{\text{CxSM}} = (R_{i2})^2 + (R_{i3})^2 \;,
\eeq
with the matrix $R$ defined in Eq.~(\ref{eq:rotsinglet}).
In the N2HDM, the singlet admixture is given by
\beq
\Sigma_i^{\text{N2HDM}} = (R_{i3})^2 \;,
\eeq
where $R$ has been introduced in Eq.~(\ref{eq:rotn2hdm}).
Also in the NMSSM the singlet admixture is obtained from the square of
the '$i3$' element of the mixing matrix,
\beq
\Sigma_i^{\text{NMSSM}} = ({\cal R}^S_{i3})^2 \;,
\eeq
with ${\cal R}^S$ introduced in Eq.~(\ref{eq:scalarrotation}). Note, that we use the
mixing matrix including higher order corrections as obtained from {\tt
NMSSMTools}. Finally, the pseudoscalar admixture $\Psi$ of the C2HDM is
defined as
\beq
\Psi_i^{\text{C2HDM}} = (R_{i3})^2 \;,
\eeq
with $R$ introduced in Eq.~(\ref{eq:c2hdmrot}). In the following we
drop the subscript and denote by $\Sigma$ and $\Psi$ the singlet and
pseudoscalar admixture of $h_{125}$, respectively. \s

{\bf The CxSM:} In the CxSM the rescaling of
all couplings to the SM particles by one common factor makes an
agreement of large singlet admixtures with the experimental data
impossible. The maximum allowed singlet admixture in the CxSM is given
by the lower bound on the global signal strength $\mu$ and amounts
to\footnote{We are neglecting here Higgs-to-Higgs decays, which is a
  valid approximation as substantial decays of $h_{125}$ into a pair of
lighter Higgs bosons would induce deviations in the $\mu$-values not
compatible with the experimental data any more.}
\beq
\Sigma_{\text{max}}^{\text{CxSM}} \approx 1 - \mu_{\text{min}} \approx
11\% \;.
\eeq

\begin{figure}[t!]
  \centering
\includegraphics[width=0.9\linewidth]{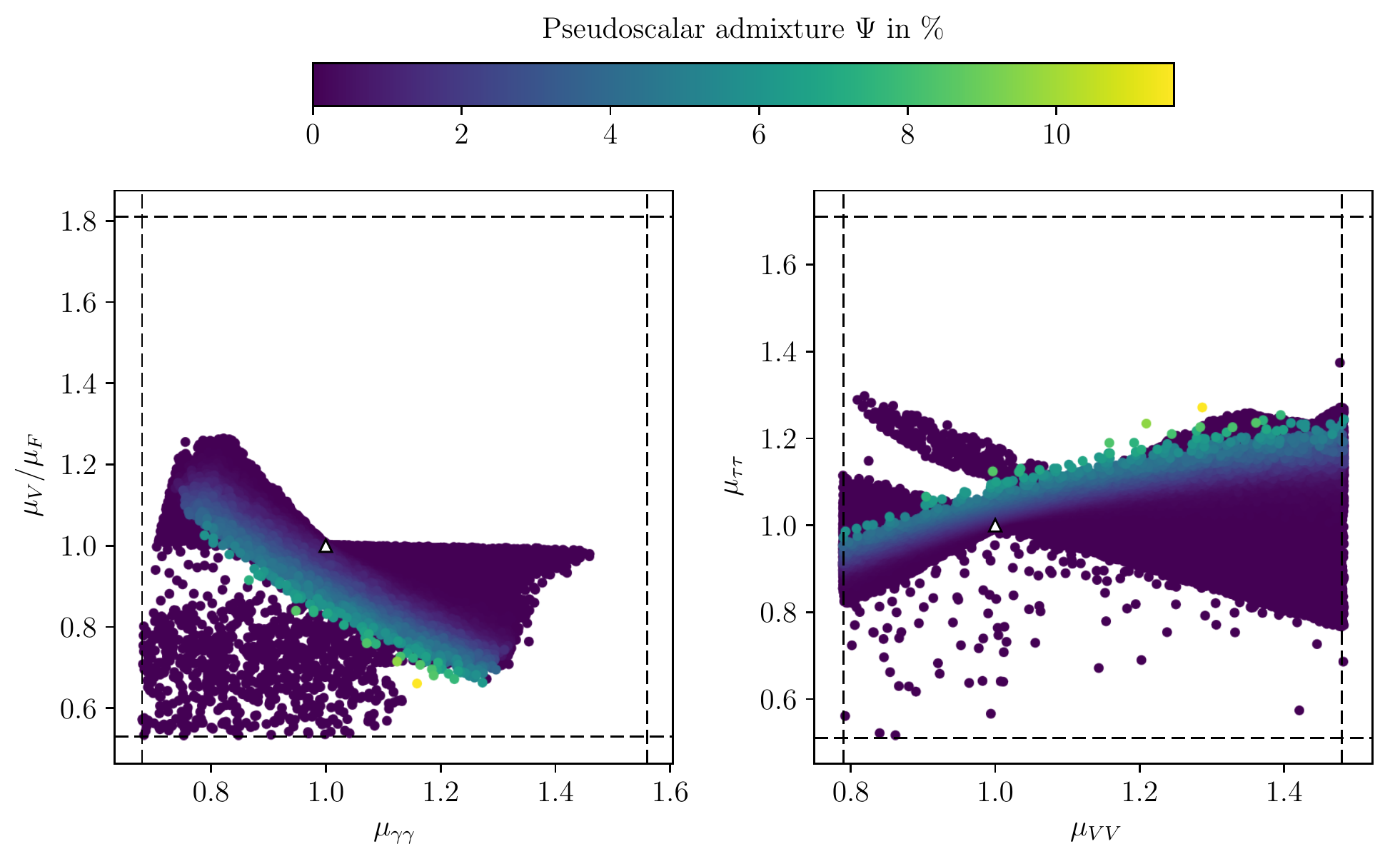}
\caption{C2HDM type II: Pseudoscalar admixture $\Psi^{\text{C2HDM}}$ of
    $h_{125}$ as a function of the most constraining signal
    strengths. The dashed lines show the experimental limits and the
    white triangle denotes the SM value.}\label{fig:C2HDM-mu}
 \end{figure}
\begin{figure}[t!]
\centering
\includegraphics[width=0.9\linewidth]{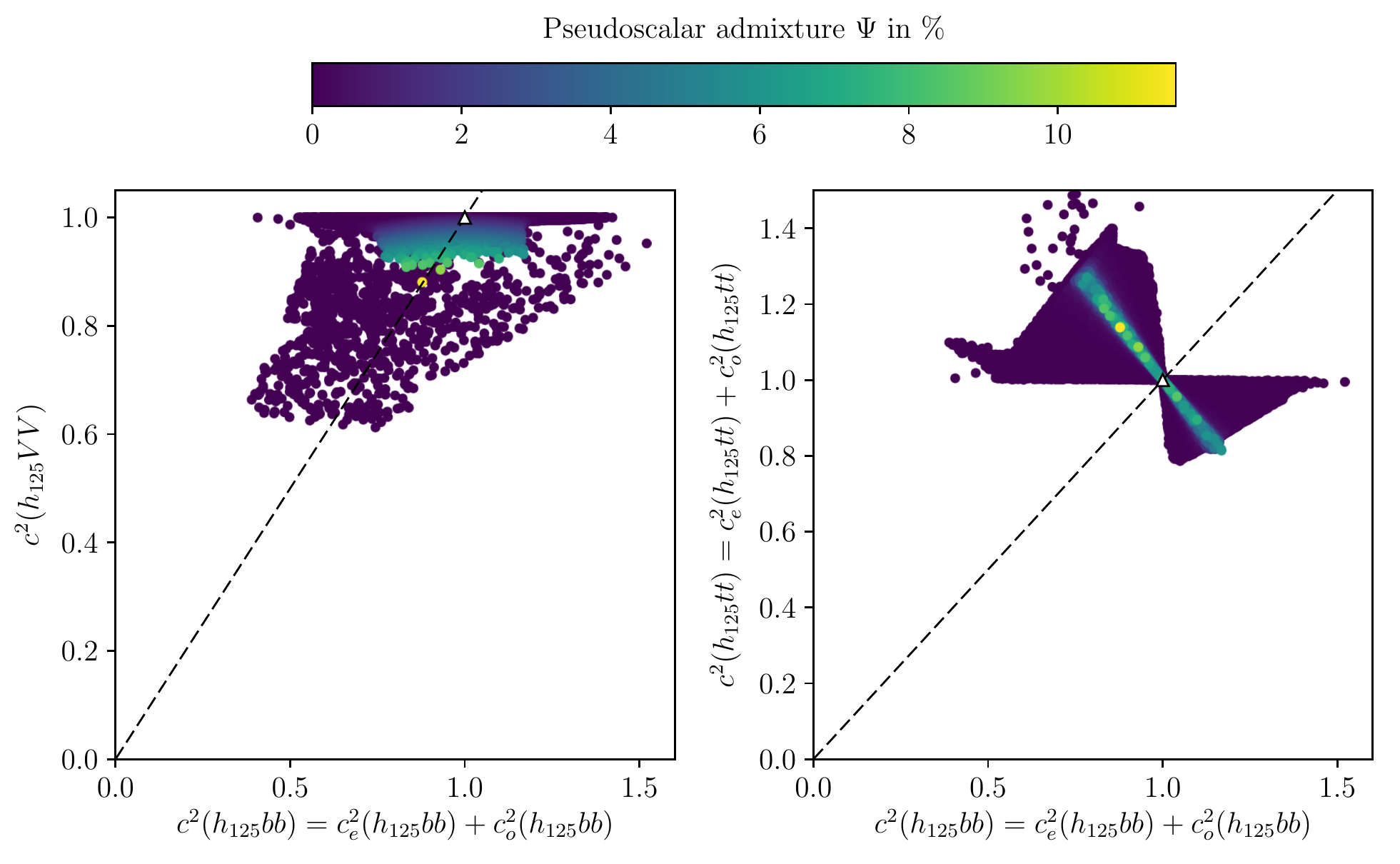}
  \caption{C2HDM type II: Pseudoscalar admixture $\Psi^{\text{C2HDM}}$ of
    $h_{125}$ as a function of the effective couplings squared. The
    white triangle denotes the SM value. The dashed line represents
    equal scaling of the couplings.}\label{fig:C2HDM-coup}
\end{figure}

{\bf The C2HDM:}
We next discuss the pseudoscalar admixture in the C2HDM. Some features
are also found in the N2HDM, so that
we do not need to repeat in detail the discussion of the N2HDM, for
which we refer to \cite{Muhlleitner:2016mzt}. We start with the
C2HDM type II. As can be inferred from
Fig.~\ref{fig:C2HDM-mu}, which shows the pseudoscalar admixture of the
C2HDM SM-like Higgs boson as a function of the most constraining signal
strengths, the pseudoscalar admixture can at most be 10\%. This is not a
consequence of the measured properties of the $h_{125}$ but due
to the restrictive bounds on the electron EDM. Without EDM
constraints 20\% would be allowed.\footnote{For a
    detailed investigation of
the C2HDM, including the analysis of the effects of EDM
constraints, we refer to a forthcoming publication
\cite{c2hdmpaper}.} Because of the rather small $\Psi$, the
properties of the $h_{125}$ in the C2HDM are well approximated by the
real 2HDM. In this limit, there are only two non-zero mixing matrix
elements that contribute to $h_{125}$. The orthogonality of the mixing
matrix leads to the sharp edges of the allowed regions visible in the
plots. In Fig.~\ref{fig:C2HDM-mu} we observe three regions of
enhanced $\mu_{\tau\tau}$ in case of small pseudoscalar admixture
(dark blue points).
One of the dark blue enhanced regions resides in the wrong-sign
limit \footnote{The wrong sign limit is the limit where the Yukawa
couplings have the relative sign to the Higgs coupling to massive gauge bosons opposite
to the SM one (see \cite{Ferreira:2014naa} for details).}
 and corresponds to the points deviating from the bulk (towards
the top left in the right and towards the bottom left in
the left plot of Fig.~\ref{fig:C2HDM-mu}).\footnote{The disconnected
  points for lower $\mu_{\tau\tau}$ values in the right plot arise from
the possibility of substantial $h_{125}$ decays into a pair of lighter Higgs
bosons. This is partly also the reason for the disconnected points in
the bottom left region of the left plot.}
Additionally, enhanced $\mu_{\tau\tau}$ rates can
be observed for non-vanishing larger pseudoscalar admixture.
This behaviour can be understood by investigating the couplings to
gauge bosons and fermions individually.
In Fig.~\ref{fig:C2HDM-coup} (left) $c^2 (h_{125}VV)$ is plotted
against $c^2(h_{125} b\bar{b}) \equiv
(c^e)^2+(c^o)^2$, with $c^{e,o}$ defined in
Eq.~(\ref{eq:yuklag}). Note that in the 2HDM type II the tree-level
couplings to down-type quarks and leptons are the same.
The right figure shows $c^2 (h_{125} t\bar{t})$
versus $c^2 (h_{125} b\bar{b})$.
The colour code indicates the pseudoscalar admixture.
While the pseudoscalar admixture reduces the couplings to gauge
  bosons the couplings to fermions can be reduced or enhanced
  irrespective of the value of $\Psi$. The enhanced rates
  are due to enhanced couplings to top-quarks, thus increasing the
  production cross section.  The additional reduction in $c(h_{125} VV)^2 = \mu_V$
leads to the reduced $\mu_V/\mu_F$, observed
for the points with larger pseudoscalar admixture in Fig.~\ref{fig:C2HDM-mu}
(left). Here we also see points with strongly reduced $\mu_V/\mu_F$
and vanishing pseudoscalar admixture. As mentioned above, these are
points residing either in the wrong-sign regime with strongly reduced
couplings to the massive gauge bosons or in the region where
substantial Higgs-to-Higgs decays of the $h_{125}$ are possible. They
are almost exclusively points of the real 2HDM.
The most enhanced $\mu_{\tau\tau}$ of up to
30\% is obtained for simultaneously enhanced
$\mu_{VV}$. It is due to the enhanced production mechanism resulting from enhanced
couplings to the top quarks in this region, as we explicitly verified,
while the involved decays remain SM-like.
The second enhanced region in the CP-conserving limit, the one in the
wrong sign regime, is due to reduced couplings to gauge bosons and
simultaneously enhanced couplings to bottom quarks. The resulting
reduced decay into $VV$ increases the branching ratio into $\tau\tau$
and thus the rate in this final state.
The third region with enhanced $\mu_{\tau\tau}$ and reduced $\mu_{VV}$
arises from enhanced
effective couplings to $\tau$ leptons and $b$-quarks. Combining this
with the fact that the couplings to massive gauge bosons cannot exceed one, the
overall branching ratio into $\tau$ pairs is enhanced.  \s
\begin{figure}[t!]
\centering
\includegraphics[width=0.9\linewidth]{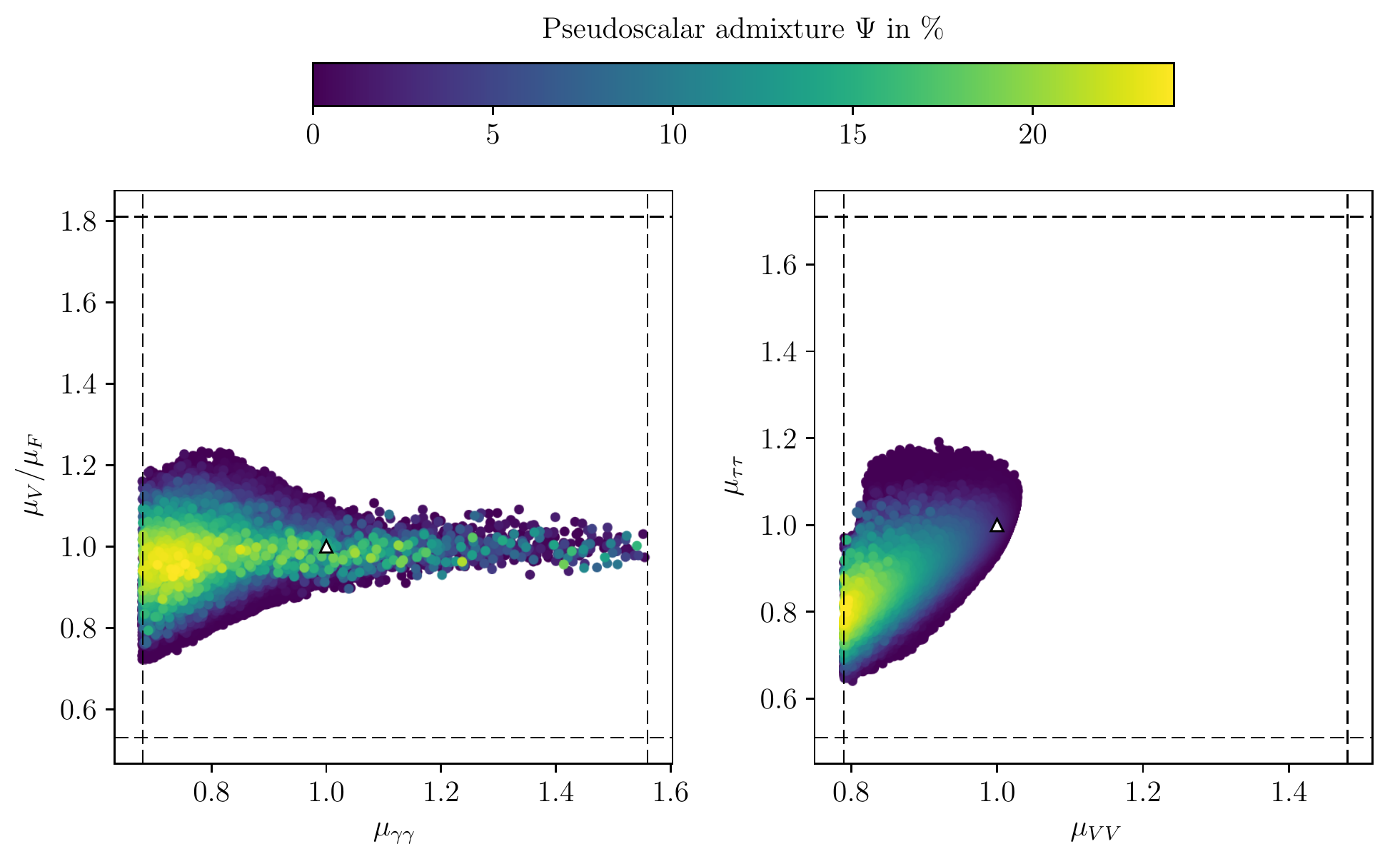}
\caption{C2HDM type I: Pseudoscalar admixture $\Psi^{\text{C2HDM}}$ of
    $h_{125}$ as a function of the most constraining signal
    strengths. The dashed lines show the experimental limits and the
    white triangle denotes the SM value.}\label{fig:t1C2HDM-mu}
 \end{figure}
\begin{figure}[t!]
\centering
\includegraphics[width=0.55\linewidth]{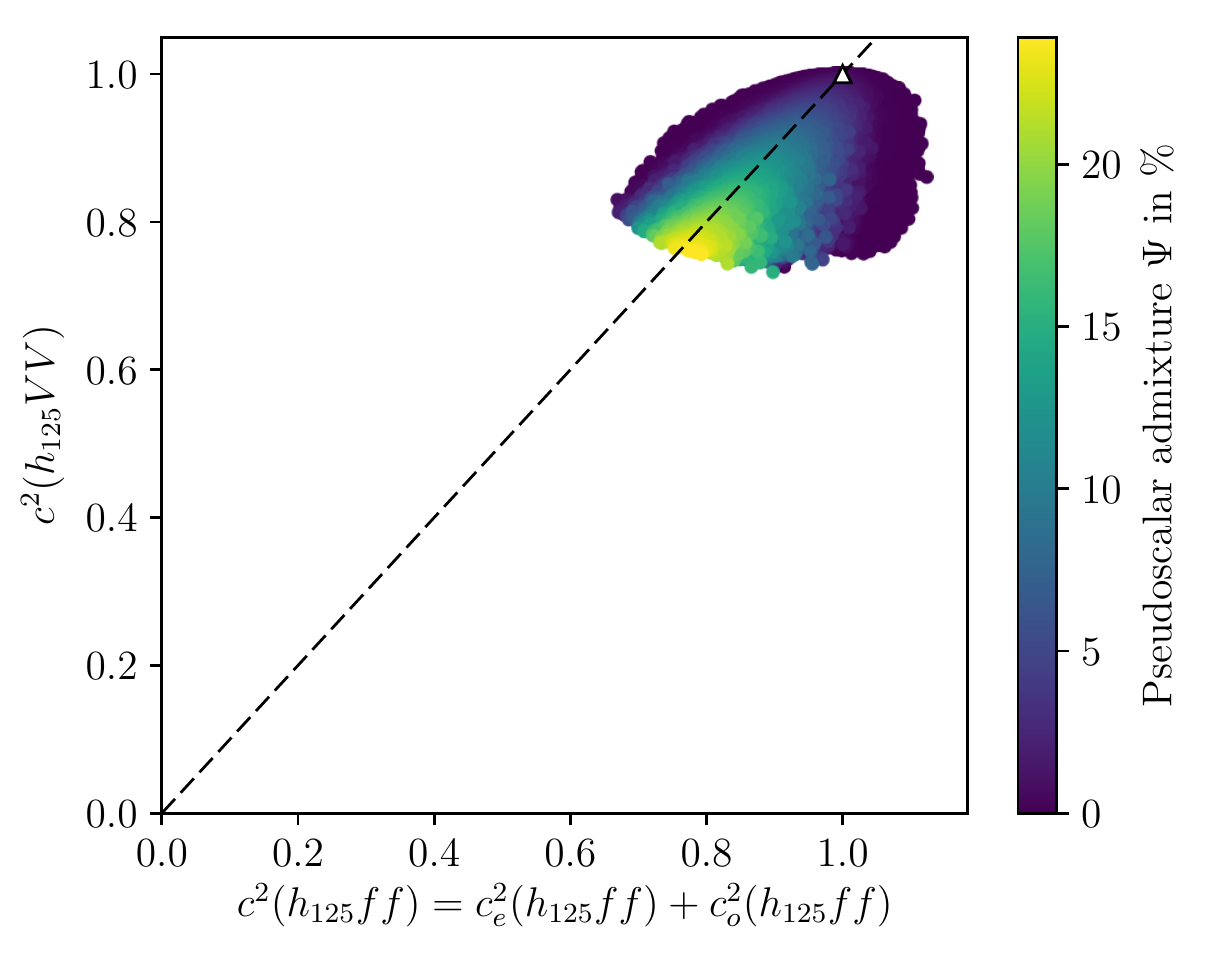}
\caption{C2HDM type I: Pseudoscalar admixture $\Psi^{\text{C2HDM}}$ of
    $h_{125}$ as a function of the effective couplings squared. The
    white triangle denotes the SM value. The dashed line represents
    equal scaling of the couplings. }\label{fig:t1C2HDM-coup}
 \end{figure}

With values of up to 25\%, {\it cf.}~Fig.~\ref{fig:t1C2HDM-mu}, larger
pseudoscalar admixtures are allowed in the
C2HDM type I compared to the type
II. This upper bound of $\Psi$ is barely affected by the EDMs
which are less constraining in the type I model. As can be inferred from
Fig.~\ref{fig:t1C2HDM-mu}, the upper
bound of $\mu_{VV}$ as well as the boundaries of $\mu_{\tau\tau}$ and
$\mu_V/\mu_F$ obtained from the combination of all the constraints in
our scan are already well inside the upper bound restrictions set by
the LHC data on these signal rates. In contrast to type II no enhanced
rates can be observed for
non-vanishing pseudoscalar admixture. The highest pseudoscalar
admixtures entail reduced signal strengths, while simultaneously the
ratio $\mu_V/\mu_F \approx 1$. In Fig.~\ref{fig:t1C2HDM-coup} $c^2
(h_{125}VV)$ is plotted against $c^2(h_{125} t\bar{t})=c^2(h_{125}
b\bar{b})$. The colour code shows that both effective couplings are
reduced almost in parallel with increasing $\Psi$, implying
$\mu_V/\mu_F \approx 1$ for large pseudoscalar admixture. We find that
for a measurement of $\mu_{\tau\tau}$ within 5\% of the SM value
pseudoscalar admixtures above 15\% are excluded. If $\mu_{VV}$ is determined
within 5\% of the SM value, $\Psi$ is even constrained to values below
7\%. In type II, only a simultaneous measurement of all $\mu$ values
within 5\% of their SM values constrain $\Psi$ to below about 3\%. \s

{\bf The N2HDM:}
In the N2HDM, the large number of free parameters allows for
significant non-SM properties of the $h_{125}$. We have investigated the singlet
admixture of the SM-like N2HDM Higgs boson in great detail in
\cite{Muhlleitner:2016mzt} and found that in the N2HDM type II singlet
admixtures of up to
55\% are still compatible with the LHC Higgs data. Interestingly, the
most constraining power on $\Sigma^{\text{N2HDM}}$ does not arise from
the best measured signal rates $\mu_{VV}$ and $\mu_{\gamma\gamma}$
which for SM-like rates in these channels still allow for singlet admixtures
of up to 50\% and 40\%, respectively. However, a measurement of
$\mu_{\tau\tau} \approx 1$ constrains $\Sigma^{\text{N2HDM}}$ to
values below about 25\%, and $\mu_V/\mu_F \ge 1$ restrict it to below
20\%. This can be understood by inspecting the involved couplings and
is due to a stronger reduction of the coupling to bottom quarks with
rising singlet admixture than the ones to top quarks and $VV$. For
details, we refer the reader to \cite{Muhlleitner:2016mzt}. Since the
N2HDM and the C2HDM coincide in their scalar sector in the limit of
vanishing singlet admixture and pseudoscalar admixture, respectively,
we observe the same enhanced regions of $\mu_{\tau\tau}$ in the limit
of the real 2HDM (type II). Away from this limit both models differ:
While non-vanishing pseudoscalar admixture allows for enhanced
$\mu_{\tau\tau}$, the singlet admixture in the N2HDM always reduces
the rates, in contrast to the C2HDM case the couplings to fermions
become smaller with rising $\Sigma$. \s

In the N2HDM type I due to the restriction
of the up- and down-type quark couplings to the same doublet we found
that the maximum allowed singlet admixture is 25\%, inducing reduced
signal strengths with simultaneously $\mu_V/\mu_F \approx
1$. The distribution of the couplings in the parameter space is
similar to that of the C2HDM type I, {\it
  cf.}~\cite{Muhlleitner:2016mzt} for comparison. Like in the C2HDM type I, the singlet
admixture is most effectively constrained, down to about 7.5\%, by a
5\% measurement of $\mu_{VV}$, while in type II $\mu_{\tau\tau}$
restricts $\Sigma$ to below 37\% (20\%) for small
(medium) $\tan\beta$ values if it is measured to 5\% within the SM value. \s

\begin{figure}[t!]
\centering
\includegraphics[width=0.9\linewidth]{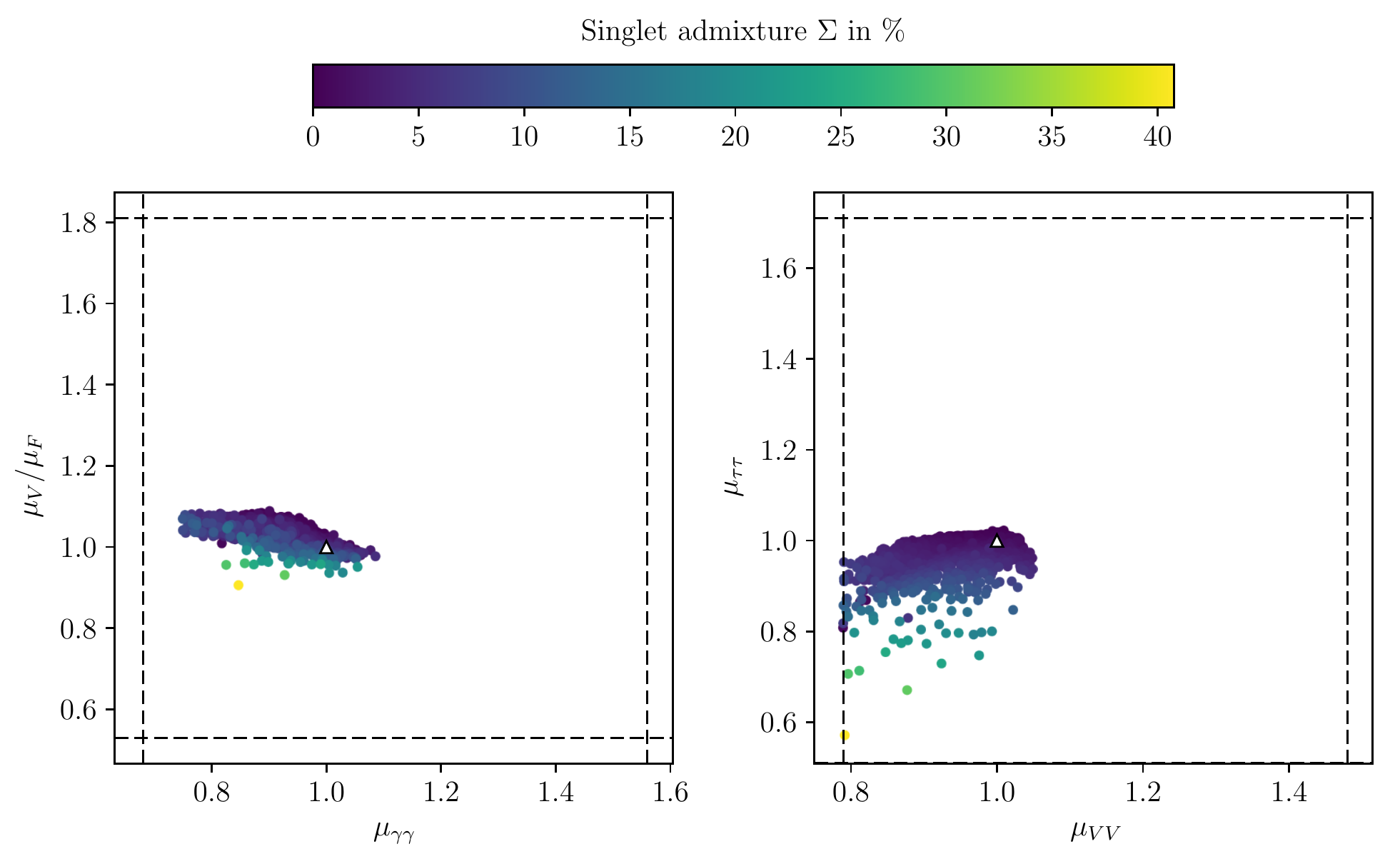}
\caption{NMSSM: Singlet admixture $\Sigma^{\text{NMSSM}}$ of
    $h_{125}$ as a function of the most constraining signal
    strengths. The dashed lines show the experimental limits and the
    white triangle denotes the
    SM value.}
\label{fig:NMSSM-mu}
\end{figure}

\begin{figure}[t!]
\centering
\includegraphics[width=\linewidth]{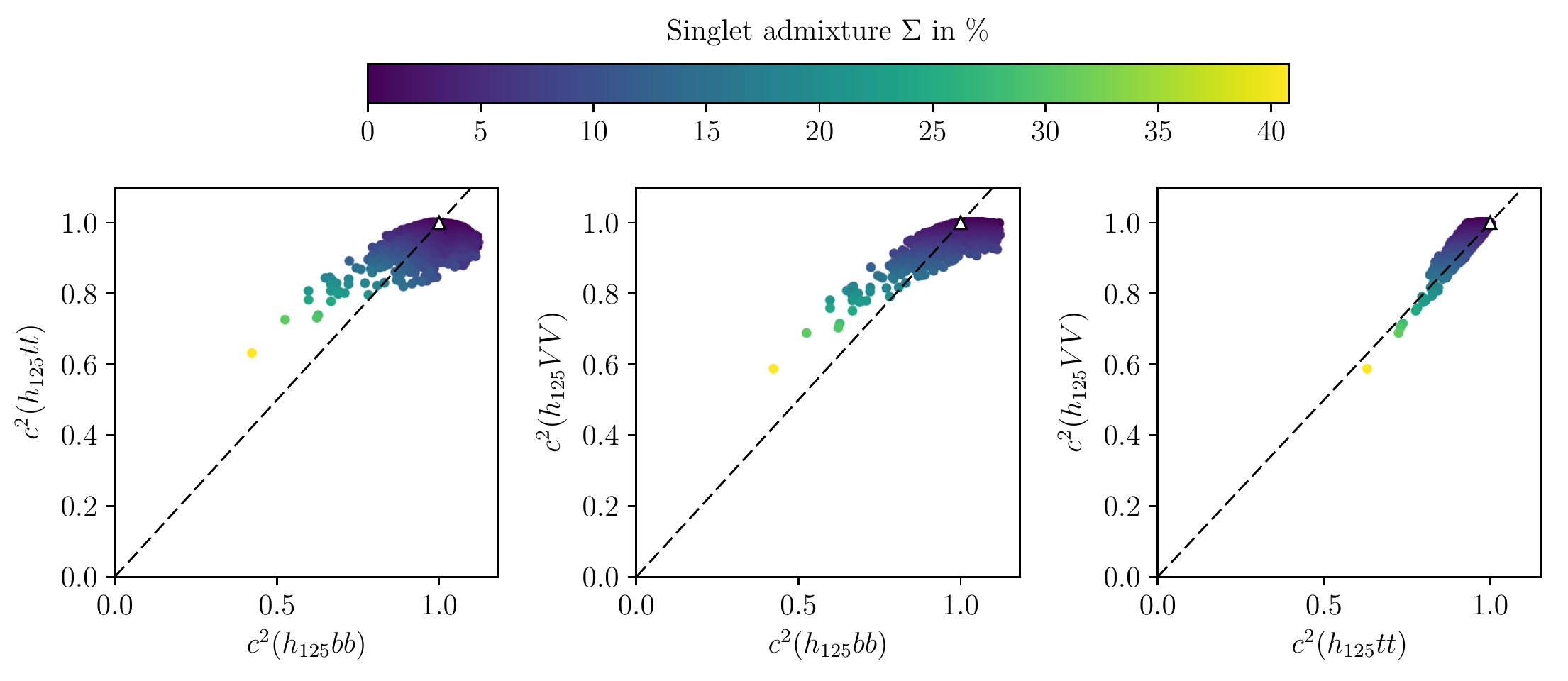}
\caption{NMSSM: Singlet admixture $\Sigma^{\text{NMSSM}}$ of the
    $h_{125}$ as a function of the effective couplings squared. The
    white triangle denotes the SM value. The dashed lines represent
    equal scaling of the couplings.}\label{fig:NMSSM-coup}
\end{figure}

{\bf The NMSSM:}
Figure~\ref{fig:NMSSM-mu} displays the singlet admixture of the NMSSM
SM-like Higgs boson as a function of the most constraining signal
strengths. These are in the left plot $\mu_{V}/\mu_F$ versus
$\mu_{\gamma\gamma}$ and $\mu_{\tau\tau}$ versus
$\mu_{VV}$ in the right one. The colour code quantifies the singlet admixture.
Due to the correlations enforced on the Higgs
sector from supersymmetry, the NMSSM parameter space is much more
constrained than the one of the N2HDM ({\it
  cf.}~\cite{Muhlleitner:2016mzt} for the corresponding plots of the
N2HDM). Furthermore, $\mu_{\tau\tau}$ cannot be enhanced by more than
a few percent, in contrast to the type II\footnote{Since
  in the NMSSM the Higgs doublets couple as in the type II Yukawa sector,
one has to compare to this type.} N2HDM, where
enhancements of up to 40\% are still
compatible with the Higgs data. The reasons for
possible (large) enhancement of $\mu_{\tau\tau}$ in the N2HDM (or
C2HDM) are all absent in the NMSSM: In the NMSSM the effective coupling to top quarks
cannot exceed 1, {\it i.e.}
\beq
c^2  (h_{125} t\bar{t}) \le 1\;,
\eeq
with $c(h_{125} t\bar{t})$ denoting the coupling modification factor
with respect to the SM coupling. This can be inferred from
Fig.~\ref{fig:NMSSM-coup}, which shows the correlations between the
NMSSM effective couplings squared together with the singlet
admixture. In the N2HDM on the other hand, the
squared top-Yukawa coupling, which controls the dominant gluon fusion
production mechanism, can be enhanced by more than 60\%. In the N2HDM
the wrong-sign regime also allows for increased $\mu_{\tau\tau}$
whereas in the NMSSM we did not find such points. Finally, the $h_{125}$
coupling squared to bottom quarks can be enhanced by more than 40\% in the N2HDM
compared to only about 15\% in the NMSSM, {\it
  cf.}~Fig.~\ref{fig:NMSSM-coup}.  \s

While in the N2HDM the ratio $\mu_V/\mu_F$ reaches its lower
experimental bound of 0.54 for $\mu_{\gamma\gamma}$
up to 1.2, {\it cf.}~\cite{Muhlleitner:2016mzt}, in
the NMSSM this ratio does not
  drop much below 1. The reason is the correlation
\beq
c^2 (h_{125} t\bar{t}) \approx c^2 (h_{125} VV) \;,
\eeq
increasing with rising singlet admixture,
as can be inferred from Fig.~\ref{fig:NMSSM-coup} (right).
The coupling to top
quarks controls gluon fusion and thus $\mu_F$, while $c^2 (h_{125} VV)
\approx \mu_V$, so that $\mu_V \approx \mu_F$. This is a consequence
of the SUSY relations together with the requirement of the $h_{125}$
to behave SM-like. 
\s

The NMSSM can still accommodate a
considerable singlet admixture of up to $\Sigma^{\text{NMSSM}} = 42$\%. Like in the N2HDM, with rising $\Sigma$ the effective coupling squared $c^2 (h_{125} b\bar{b})$ is reduced more strongly than $c^2(h_{125} VV)$ and $c^2 (h_{125} t\bar{t})$, as can be inferred from
Fig.~\ref{fig:NMSSM-coup} (left and middle). The enhancement in the branching ratios due
to the reduced dominant decay into $b\bar{b}$ and hence the smaller
total width is large enough to counterbalance the reduction in the
production. The coupling strength to $\tau$'s is reduced in the same
way as the one to bottom quarks when the singlet
admixture increases. As there are no other means to enhance
$\mu_{\tau\tau}$ in order to compensate for the effects of non-zero singlet admixture,
the $\mu_{\tau\tau}$ is very constraining and even more constraining
than in the N2HDM. A measurement of
$\mu_{\tau\tau}$ within 5\% of the SM value would exclude singlet
admixtures larger than 8\%.  \s

\section{Signal Rates of the non-SM-like Higgs Bosons \label{sec:sigrates}}
In this section we show and compare the rates of all neutral
non-SM-like Higgs bosons in the
most important SM final state channels.
Assuming that in a first stage of discovery only one additional Higgs
boson besides the $h_{125}$ has been discovered we also investigate the
question if in this situation, {\it i.e.}~before the discovery of further Higgs
bosons, we are already able to distinguish between the four models
discussed here. As the determination of the CP properties of the new
Higgs boson is not immediate and takes some time to accumulate a
sufficiently large amount of data, we assume that the CP
properties of the second discovered Higgs boson are not known, so that we have to treat
the CP-even, CP-odd and CP-mixed (in the C2HDM) Higgs bosons of our
models on equal footing.
Again, we denote by $H_\downarrow$ the lighter and by
$H_\uparrow$ the heavier of the two neutral non-$h_{125}$ CP-even
or CP-mixed (for the C2HDM) Higgs bosons. The pseudoscalar of the
N2HDM is denoted by $A$ and the two pseudoscalars of the NMSSM by
$A_1$ and $A_2$, where by definition $m_{A_1} < m_{A_2}$.
The signal rates that we show have been obtained by
multiplying the production cross section with the corresponding
branching ratio obtained from {\tt sHDECAY}, {\tt
  N2HDECAY}, {\tt NMSSMCALC} and a private version
including the CP-violating 2HDM (to
be published in a forthcoming paper). For the production we use
\beq
\sigma_{\Phi} = \sigma_{\Phi} (ggF)
+ \sigma_{\Phi} (bbF) \;,
\eeq
computed for a c.m.~energy of $\sqrt{s}=13$~TeV with {\tt SusHi} at
NNLO QCD using the effective $t$ and $b$ couplings of the respective
model. Here $\Phi$ generically denotes any of the CP-even, CP-odd and
CP-mixed neutral Higgs bosons of our models.
Production through bottom-quark fusion is included in order to
account for possible large $b$-quark couplings. None of our models
can lead to enhanced couplings to vector bosons, so that we neglect
the sub-leading production through vector boson fusion. As
Higgs-strahlung and associated production are negligible compared to
$ggF$ and $bbF$, we neglect these production channels as well. Furthermore, for all rates we
impose a lower limit of 0.1~fb. \s

\begin{figure}[ht!]
\centering
\includegraphics[width=0.9\linewidth]{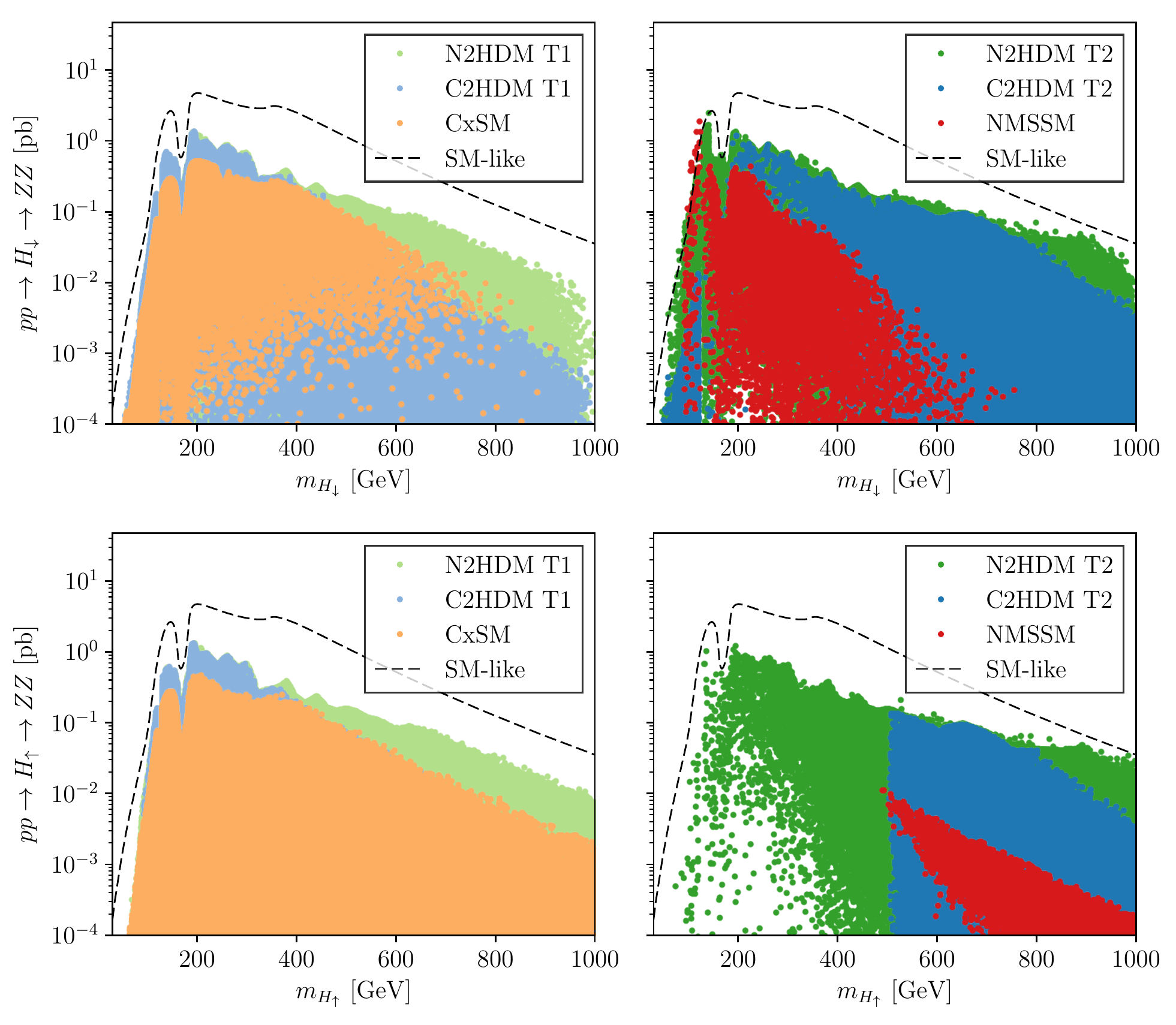}
\caption{Signal rates for the production of $H_\downarrow$ (upper) and
  $H_\uparrow$ (lower) decaying
  into a pair of $Z$ bosons at the $\sqrt{s}=13$~TeV LHC as a function
  of its mass. Left: the CxSM (orange), the type I N2HDM
  (fair green) and C2HDM (fair blue). Right: the NMSSM (red) and the
  type II N2HDM (dark green) and C2HDM (dark blue). The
  dashed black line denotes the signal rate
  of a SM Higgs boson of the same mass.}\label{fig:Hd-ZZ}
\end{figure}

{\bf Signal Rates into $ZZ$:}
In Fig.~\ref{fig:Hd-ZZ} we depict the signal rates into $ZZ$.
The rates of the two non-SM-like Higgs bosons of the C2HDM are shown
together with the CP-even Higgs bosons of the other models in one
plot, although they can have a more or less important pseudoscalar
admixture. Note also, that there are no rates for the pure pseudoscalar
Higgs bosons of the N2HDM and NMSSM, as they do not couple to massive gauge
bosons at tree-level. For all our models the sum rule
\beq
\sum_{i=1}^3 c^2 (H_i VV) = 1 \label{eq:sumrulevv}
\eeq
for the CP-even and C2HDM CP-mixed Higgs bosons holds, imposed by
unitarity constraints. Since the $h_{125}$ requires
substantial couplings to gauge bosons in order to comply with the
experimental results in the $ZZ$ and $W^+ W^-$ final states, the sum
rule forces the gauge coupling of $H_\downarrow$ (and also
$H_\uparrow$) to be considerably below the SM value. The room for
deviations of the $h_{125}$-Higgs boson coupling to gauge bosons from
the SM value mainly depends on the number of free parameters of the
model that can be used to accommodate independent coupling
variations. This allows {\it e.g.}~a reduction of the
decay width into gauge bosons to be compensated by the reduction
of the total width and/or an increase in the production cross
section. \s

In the CxSM the common scaling of all Higgs couplings combined with
the sum rule Eq.~(\ref{eq:sumrulevv}) and
the fact that experimental data allow for $\mu_{h_{125}}$ down to
about 0.9 enforces
\beq
c^2_{\text{CxSM}} (H_{\downarrow/\uparrow} VV) \lsim 0.1 \;.
\eeq
As all CxSM Higgs couplings are reduced compared to the SM the
production cross sections cannot be enhanced in this model, so that
altogether not only the rate into $VV$ but all CxSM rates are below
the SM reference in the whole $m_{H_{\downarrow/\uparrow}}$ mass range
so that the discovery of additional Higgs bosons in the CxSM may
proceed through Higgs-to-Higgs decays \cite{Costa:2015llh}.
\s

Also for the remaining models overall we observe reduced rates
compared to what would be expected in the SM for a Higgs boson of the same
mass, except for the low-mass region. The resulting rates are a
combination of sum rules and the behaviour of the
Yukawa couplings.\footnote{In the NMSSM additional squark
  contributions in the dominant gluon fusion production cross section
  or stop, chargino and charged Higgs contributions in the loop decay
  into photons play a role if the loop particle masses are light
  enough \cite{King:2012tr}.} As the $h_{125}$ takes a large portion of the
coupling to gauge bosons, the $H_{\downarrow/\uparrow} VV$ coupling necessarily
cannot be substantial. Models with more parameters, however, like
the ones discussed here, allow for larger deviations of the
$h_{125}$ couplings from the SM expectations. This allows the remaining Higgs bosons
to have larger couplings, while maintaining compatibility with any coupling sum rules.
We will discuss the implications of such sum rules in great detail in the next section. As we have seen before the couplings to fermions can also be
enhanced in some models. Finally, due to SUSY
relations the NMSSM has less freedom than the N2HDM. Overall the
combination of all these effects leads to the rates in most mass regions being largest in the N2HDM. Furthermore, the
rates in the type I models are (somewhat) smaller than in the
corresponding type II models, as
in the former we have the additional constraint that the up- and
down-type couplings cannot be varied independently. \s

The behaviour of
the NMSSM cross sections can be best understood by looking at the
nature of the Higgs boson under investigation. This is summarized in
Table~2 of Ref.~\cite{King:2014xwa}. The $H_\downarrow$ with mass
below 125~GeV behaves singlet-like but can become doublet-like in
regions with strong doublet-singlet mixing which happens in mass
regions close to 125~GeV. This is why here the rates can become
SM-like or even exceed the SM reference value. In this case, where the
second-lightest Higgs $H_2 \equiv h_{125}$, the heaviest one, $H_\uparrow$,
is doublet-like. In case the lightest Higgs boson $H_1 \equiv
h_{125}$, $H_\downarrow$ is singlet (doublet)-like
for small (large) $\tan\beta$, and $H_\uparrow$ takes the opposite
role. Despite the fact that for masses above 125 GeV either $H_\downarrow$ or
$H_\uparrow$ are doublet-like their couplings to massive gauge bosons are
suppressed as discussed in \cite{King:2014xwa} so that the NMSSM rates
always remain well below the SM reference. \s

Since all the rates of the various models overlap, a distinction based
on this criterion is difficult. One can state, however, that
an observation of a neutral scalar with an ${\cal O} (100\mbox{ fb})$ rate in the $ZZ$
  channel for a mass $\gsim 380$~GeV may be sufficient to exclude
  the NMSSM. Furthermore
the observation of rates of 30-50~fb in the high mass region between 800
  and 1000~GeV can only be due to the N2HDM (type II), within our set
  of models. This region is being tested by the experiments, which are
  due to achieve soon the luminosity necessary to probe such high
  rates \cite{CMS:2017sbi}. \s

\begin{figure}[t!]
\centering
\includegraphics[width=0.9\linewidth]{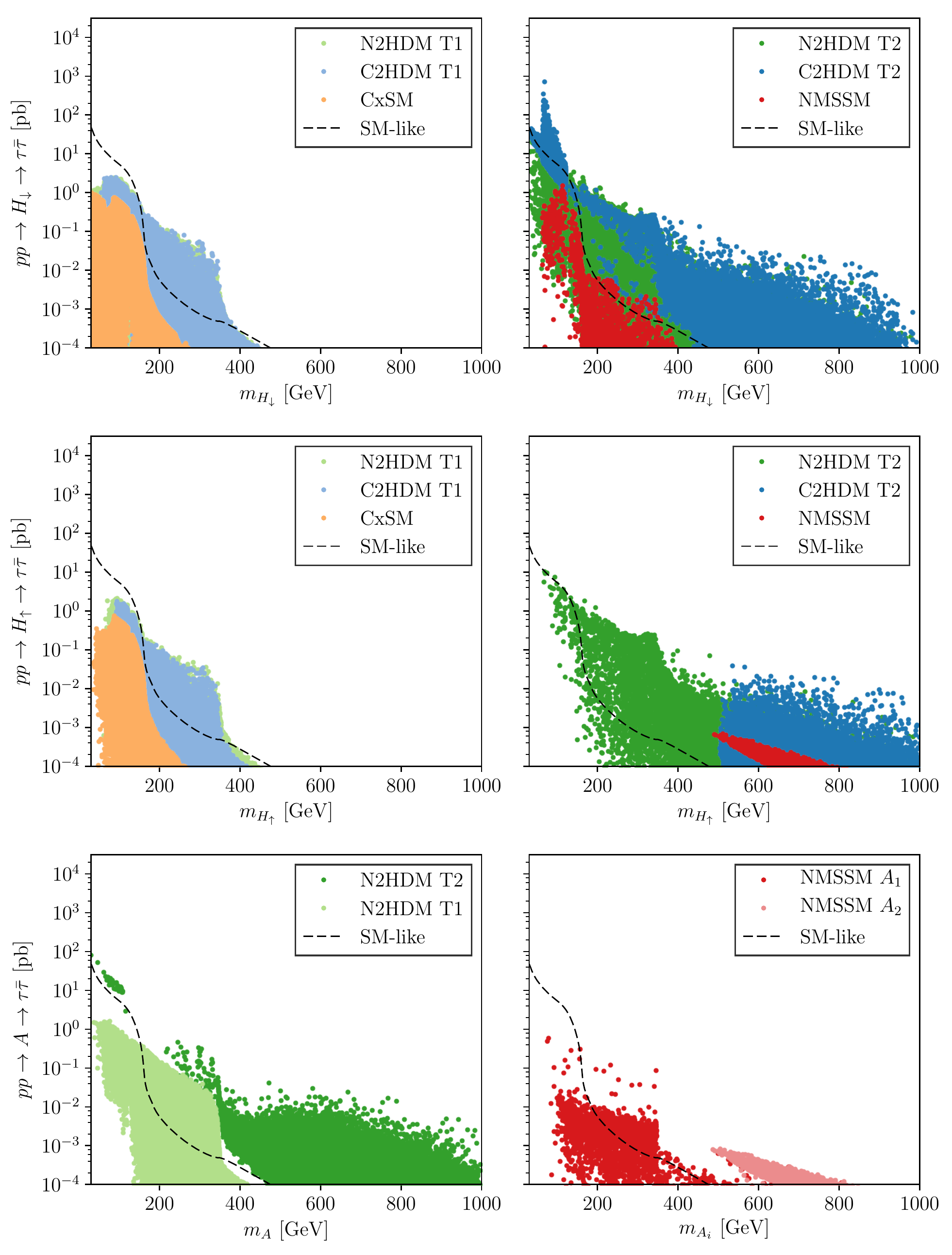}
\caption{Same as Fig.~\ref{fig:Hd-ZZ} but for the $\tau$-pair final
  state. Also included, in the lower row: tauonic decays of the N2HDM
  pseudoscalar $A$ (left) for the types I (fair green) and II (dark
  green) and of the NMSSM pseudoscalars (right) $A_1$ (red) and $A_2$
  (rose).}
\label{fig:Hd-tautau}
\end{figure}
\begin{figure}[t!]
\centering
\includegraphics[width=0.9\linewidth]{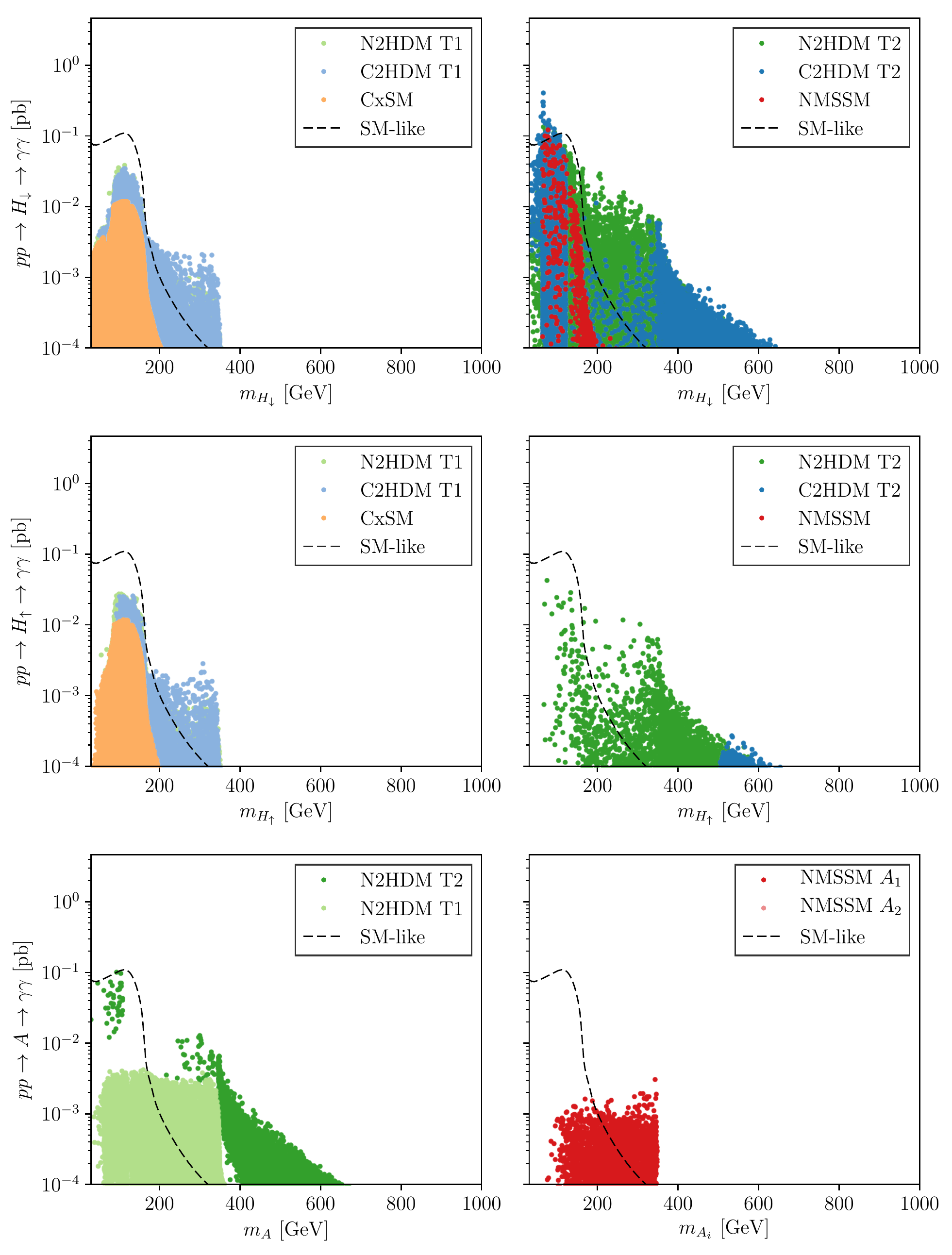}
\caption{Same as Fig.~\ref{fig:Hd-tautau} but for the photonic final
  state.}\label{fig:Hd-gamgam}
\end{figure}
{\bf Signal Rates into $\tau\tau$:}
Figure~\ref{fig:Hd-tautau} displays the signal rates into the
$\tau$-pair final state for the various models. In all models apart
from the CxSM the $H_{\downarrow/\uparrow}$ couplings to $\tau$-pairs can be
enhanced above the SM value, so that enhanced rates are possible
provided the production cross section is not too strongly
suppressed. In particular in the C2HDM, the incoherent addition of the
scalar and pseudoscalar contributions to both the $ggF/bbF$
production and the partial width into $\tau\bar{\tau}$ leads to
enhanced rates. This concerns the points with non-vanishing $\Psi$
where $m_{H_{\uparrow/\downarrow}} \gsim 500$~GeV. All other points reside
in the limit of the CP-conserving 2HDM, as discussed above. Note that
the points of the type II N2HDM and C2HDM (here in the $\Psi \to 0$
limit) with enhanced $\tau$ rates for $m_{H_\downarrow} \lsim 200$~GeV
are about to be constrained (or excluded) experimentally~\cite{CMS:2016rjp}. The very
enhanced points at $m_{H_\downarrow}=70-80$~GeV are due to associated
production with bottom quarks for large values of $\tan\beta$ in the real 2HDM limit of both the C2HDM and N2HDM. In this
mass region no exclusion limits exist so far so that these points are
still allowed. This should encourage the experiments to perform
analyses in this mass region. For $m_{H_\downarrow} \lsim 65$~GeV,
limits exist from the SM-like Higgs data, as $h_{125}$ can decay
off-shell into a pair of $H_\downarrow$ which could possibly spoil the
measured $\mu$-values of $h_{125}$. The
NMSSM rates are explained as follows: Irrespective of $\tan\beta$ the
$H_\downarrow$ is singlet-like for $m_{H_\downarrow} < 125$~GeV and
becomes more and more doublet-like in the vicinity of $h_{125}$ so
that its rates become more SM-like. For $m_{H_\downarrow} \ge 130$~GeV
$H_\downarrow$ is singlet-(doublet-)like for small
(large) $\tan\beta$.
The applied limits on
the SUSY masses turn out to restrict the NMSSM parameter range to
smaller values of $\tan\beta$, so that $H_\downarrow$ is singlet-like
in this mass region and its rates are below the SM reference values.
The $H_\uparrow$ is doublet-like for $\tan\beta$ small and $h_{125}$ either $H_1$ or
$H_2$. As $\tan\beta$ cannot become large, however, its rates are
not much above the values that would be obtained in the SM case.
\s

The lower two plots display the production cross sections of the N2HDM
pseudoscalar $A$ in the N2HDM (left plot) and of the two NMSSM pseudoscalars
$A_1$ and $A_2$. The SM-like Higgs limit is also included in the dashed line as a reference\footnote{Note that the production cross section for a CP-odd Higgs is larger than for a CP-even one with the same mass.}. Again in
N2HDM type II the rates are larger than in type I. In the range $130
\mbox{ GeV} \le m_A \lsim 200$~GeV there are hardly any points due to
the LHC exclusion limits \cite{CMS:2016rjp}. The enhanced rates for
$m_A \le 120$~GeV are on the border of being excluded. The shape of
the NMSSM $A_{1,2}$ distributions is again
explained by the singlet-/ doublet-nature of these particles. The
lighter of the two pseudoscalars, $A_1$, is singlet-like for $m_{A_1}
\lsim 380$~GeV. Still, in the region above the $Z$-pair and below the
top-quark pair threshold, the $\tau\tau$ rates can exceed the SM
reference, as the decay into $ZZ$ bosons which is dominant here in the
SM, is absent. The sharp edge at 350~GeV is due to the opening of the
decay into top-quarks. The $A_2$ is correspondingly doublet-, {\it
  i.e.} MSSM-like, explaining its larger rates for
  the same mass value. \s

The comparison of all models shows that it is
impossible to distinguish the models based on these rates. Only the
CxSM can be excluded if rates above the SM are found, as expected.
\s

\begin{figure}[t!]
\centering
\includegraphics[width=0.9\linewidth]{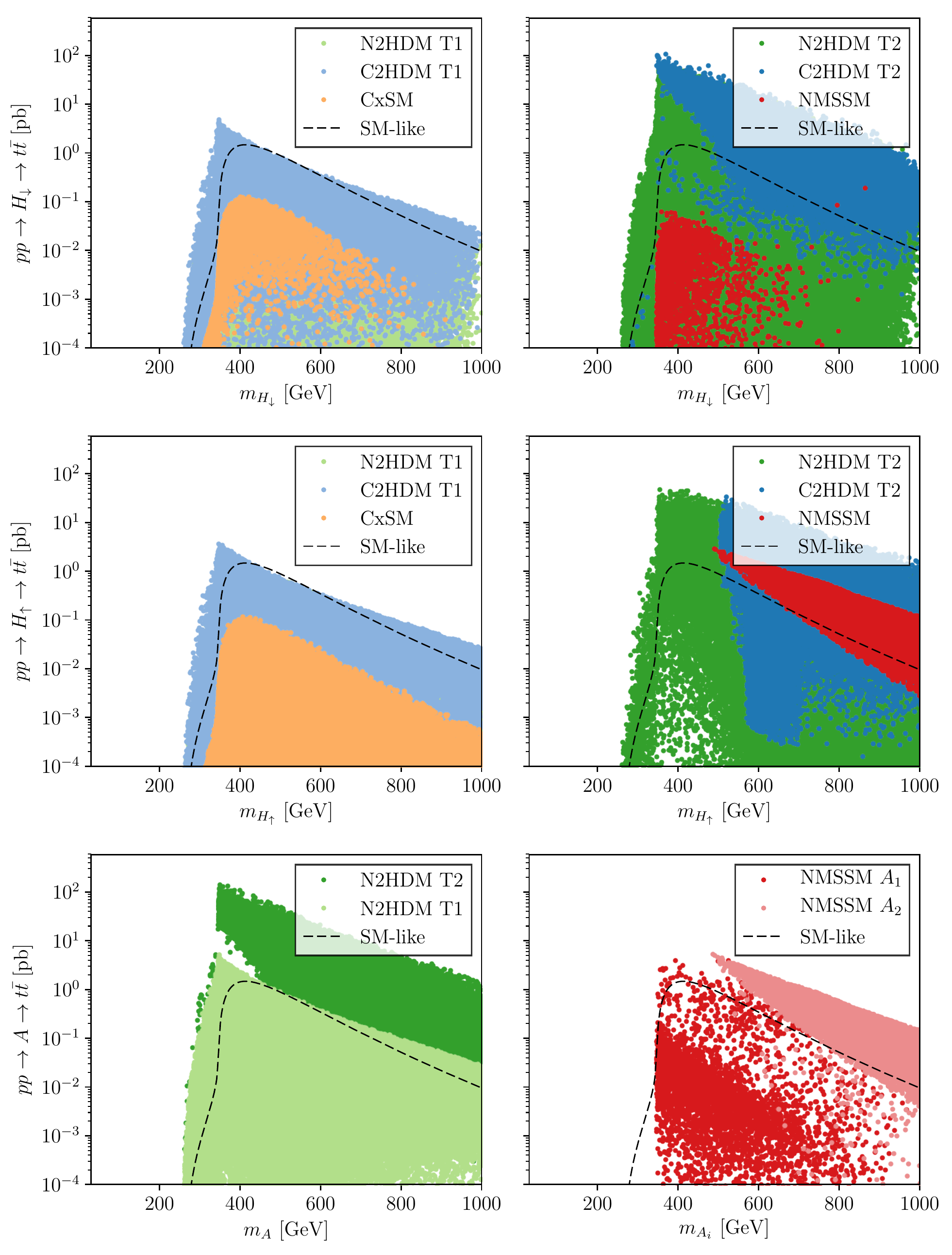}
\caption{Same as Fig.~\ref{fig:Hd-tautau} but for the top-quark pair final
  state.}\label{fig:Hd-tt}
\end{figure}
{\bf Signal rates into $\gamma\gamma$:}
In Fig.~\ref{fig:Hd-gamgam} we study the rates in the photonic final state. The distributions show the same shape as for the tauonic final
state, only moved downwards to smaller rates. Interesting are the
enhanced photonic rates for mass values below 125~GeV in the upper
right plot for the NMSSM and the type II N2HDM
and C2HDM. The latter, however, are points in the limit of the real
2HDM. The N2HDM points are hidden behind the NMSSM ones and reach
equally large rates. The even higher 2HDM points will soon be constrained (or excluded)
once the experimental analyses investigate this mass range. These
findings, however, should further encourage searches in these mass
regions in the tauonic and photonic final states. Also in the photonic final
state, the distinction of the model based on the final states is
difficult. Only the observation of rates above 5 fb in the mass range between 
130 and 350~GeV would indicate a (non-supersymmetric) extended Higgs sector 
of type II Yukawa structure as the only valid model among the ones we are discussing. 
However, these rates are experimentally challenging. \s

{\bf Signal rates into $tt$:}
Finally, in Fig.~\ref{fig:Hd-tt} the rates into top-quark pair final
states are shown. The largest rates are achieved in the type II N2HDM
and C2HDM, where the C2HDM points cover the N2HDM points, which reach
equally high rates. Note, however, that again all points below 500~GeV
are only obtained in the limit of the real 2HDM and not related to any
CP-mixing. The NMSSM $H_\downarrow$ rates are far below the SM ones,
as $H_\downarrow$ is singlet-like for small values
  $\tan \beta$. It behaves doublet-like for large values of $\tan\beta$. But then the
decay into tops is suppressed.
However, the $H_\uparrow$ is doublet-like for
small $\tan\beta$ values inducing rates above the
SM ones. Also the NMSSM pseudoscalar $A_2$ is doublet-like for small
$\tan\beta$ values, so that large rates are obtained, while $A_1$ is
doublet-like for large $\tan\beta$, so that large rates are
precluded. In the N2HDM type II small values of $\tan\beta$ are still
allowed so that large rates can be obtained for $A$, which couples
proportionally to $1/\tan\beta$ to up-type quarks both in type I and
II. With rates of up to
${\cal O} (100 \mbox{ fb})$ and more, the search for heavy
(pseudo)scalars in the top-pair final state in the 2HDM, N2DHM and
NMSSM becomes interesting. A distinction of the models is
difficult. The NMSSM, however, can be excluded if rates above 20 fb
are observed in the top-pair final state.

\section{Coupling Sums \label{sec:sumrules}}
In this section we investigate what can be learnt about the underlying
model from the coupling patterns of the discovered Higgs bosons. We
study the gauge boson sum
\beq
\Pi^{(i)}_{VV} = \sum_{j=1}^i |c(H_j VV)|^2
\eeq
and the Yukawa sum
\beq
\Pi^{(i)}_{\text{Yuk}} = \frac{1}{\sum_{j=1}^i |c(H_i \tau\bar{\tau})|^2}
+  \frac{1}{\sum_{j=1}^i |c(H_i t\bar{t})|^2} \;.
\eeq
As evident from these definitions
\beq
\Pi^{(i)}_{VV} \le \Pi^{(i+1)}_{VV} \quad \mbox{and} \quad
\Pi^{(i)}_{\text{Yuk}} \ge \Pi^{(i+1)}_{\text{Yuk}} \;. \label{eq:yuksumrule}
\eeq
The sums are performed over the CP-even Higgs bosons of the CxSM,
N2HDM and NMSSM, and over the CP-violating neutral Higgs bosons of the
C2HDM. In the C2HDM and the N2HDM, the Yukawa sum depends on the way the
Higgs doublets couple to the fermions. In type II, the
coupling to $\tau$ leptons can be exchanged by the $b$-quarks, leading
to the same result, which for the sum over all neutral Higgs bosons is
independent of $\tan\beta$. In the remaining
types, this Yukawa sum can be dependent on $\tan\beta$. In our analysis
we assume the experimental situation that only one additional neutral
Higgs boson with non-vanishing gauge coupling has been discovered.\s

Note that for the unitarity of scattering processes to be fulfilled
the couplings of the Higgs bosons to the gauge bosons and to
the fermions, respectively, have to take a specific form. All our models are weakly
interacting, and the couplings fulfil the unitarity requirement,
expressed through sum rules
\cite{Cornwall:1974km,Gunion:1990kf,Grzadkowski:1999ye}.
The specific
form of the coupling sum rules can be derived from 2-to-2
scattering processes, by requiring these to fulfil unitarity. Thus,
longitudinal gauge boson scattering into a pair of
longitudinal gauge bosons implies that $\Pi^{(i)}_{VV}$ is equal to 1 if the sum is
performed over {\it all} Higgs bosons coupling to the gauge bosons. If
one Higgs boson is missed the sum rule is violated. The sum over the
fermion couplings has not been derived from a 2-to-2 scattering
process. Instead it has been constructed such that it yields 1 for the NMSSM and
the type II N2HDM when the complete sum over all CP-even Higgs bosons
is performed. The outcome of the Yukawa sum defined in
Eq.~(\ref{eq:yuksumrule}) depends on the way the Higgs doublets couple to
the fermions, so that the sums for the N2HDM type I and the C2HDM type I and II
depend on the model parameter $\tan\beta$.
In the following we will investigate how the gauge boson and Yukawa
sums in our models change if the sum is performed only over a subset
of the Higgs bosons. 
In case not all neutral Higgs bosons of a given model are included in
the gauge boson sum, it will deviate from 1.
In the MSSM and the CP-conserving 2HDM, however,
the sum over two discovered CP-even Higgs bosons is complete and yields
1 both for the gauge boson sum rule and the Yukawa sum (2HDM type II only). \s

At the LHC the Higgs couplings can only be extracted by applying model
assumptions. The accuracy at 68\% C.L. on the $VV$ and $\tau\tau$
couplings to be expected for an integrated luminosity of 300~fb$^{-1}$
(3000~fb$^{-1}$) is about 10\% (slightly better than 10\%), on the
$t$-quark coupling about 15\% ($\sim$12\%) and around 20\% (16\%) for
the $b$-quark coupling, see {\it
  e.g.}~\cite{Lafaye:2009vr,Klute:2012pu,Plehn:2012iz,Bechtle:2014ewa}. The
model-independent coupling measurements at a linear collider (LC)
improve these precisions to a few percent at a c.m.~energy of 500~GeV
with an integrated luminosity of 500~fb$^{-1}$ \cite{Bechtle:2014ewa,Klute:2013cx,AguilarSaavedra:2001rg,Battaglia:2004mw,Djouadi:2007ik,Baer:2013cma}. The
combination of the high-luminosity LHC and LC leads to a
further improvement on the extracted accuracy. Due to the lower
statistics the precision on the Higgs couplings of the non-SM-like
Higgs bosons might be somewhat lower. Their CP-even or -odd nature can be tested in
an earlier stage after discovery by applying different spin-parity
hypotheses. The measurement of possibly CP-violating admixtures, however,
requires the accumulation of a large amount of data, so that a
dominantly CP-even Higgs boson of the C2HDM can be misinterpreted as
CP-even and is taken into account in this analysis, as also argued
above. \s

In the C2HDM and CxSM all three neutral Higgs bosons mix so that the
coupling sum analysis can straightforwardly be applied. For the N2HDM and
NMSSM, however, it has to be made sure that the additionally
discovered Higgs boson included in the sum, is CP-even. If the
observed particle is observed in the $ZZ$ decay channel, it cannot be
purely CP-odd \cite{Heinemeyer:2013tqa,deFlorian:2016spz,Fontes:2015xva}. Therefore, we require for the non-SM-like Higgs boson
\beq
ggF \to H_{\downarrow/\uparrow} \to ZZ > 10 \mbox{ fb} \;.
\label{eq:cpnature}
\eeq
This should be observable at the high-luminosity LHC, especially if properties of the particle are known from prior observations in other channels. This still allows for $H_{\downarrow/\uparrow}$ to be a CP-mixed
state, which leads to interesting phenomenological consequences for the
C2HDM. In \cite{Arhrib:2006rx, Bernreuther:2010uw, Diaz-Cruz:2014aga} it has been shown that the
loop-induced decay $A \to
ZZ$ of the pure pseudoscalar in the CP-conserving 2HDM can lead to
considerable rates. Assuming that a similar behaviour might be possible
in the N2HDM\footnote{There exists no corresponding study for the
  N2HDM so far.}, the $ZZ$ decay channel might not be sufficient to
unambiguously identify the CP nature of the Higgs boson, but
other measurements like {\it e.g.}~the angular distributions in $Z$- and
$\gamma$-pair final states or fermionic decay modes could be used to
identify the CP nature of the discovered particle, see {\it
  e.g.}~\cite{Choi:2002jk,Miller:2001bi,Godbole:2007cn,Berge:2008wi,Berge:2008dr,Berge:2011ij,Berge:2013jra,Heinemeyer:2013tqa,Berge:2014sra},
and to ensure no CP-odd particle is included in the sum. \s

With the coupling sums at hand, we want to investigate the following
questions in the next three subsections:
\begin{itemize}
\item Assuming that only two neutral CP-even (or, for the C2HDM, two dominantly
  CP-even) Higgs bosons have been found,
  can we decide based on the coupling sums if the CP-even (or, for the C2HDM,
  CP-mixed) Higgs sector is
  complete (like {\it e.g.}~in the MSSM or CP-conserving 2HDM that
  incorporate only two CP-even Higgs bosons) or if we are missing the
  discovery of the remaining Higgs bosons of an extended Higgs sector?
\item If this is possible, does the inspection of the pattern of the
  coupling sums allow us to draw conclusions on the mass scale of the
  missing Higgs boson?
\item Furthermore, can we distinguish between the various models
  investigated here on the basis of the sum distributions of the
  two discovered Higgs bosons?
\end{itemize}

\subsection{Gauge Boson Coupling Sums}
\begin{figure}[t!]
\centering
\includegraphics[width=0.9\linewidth]{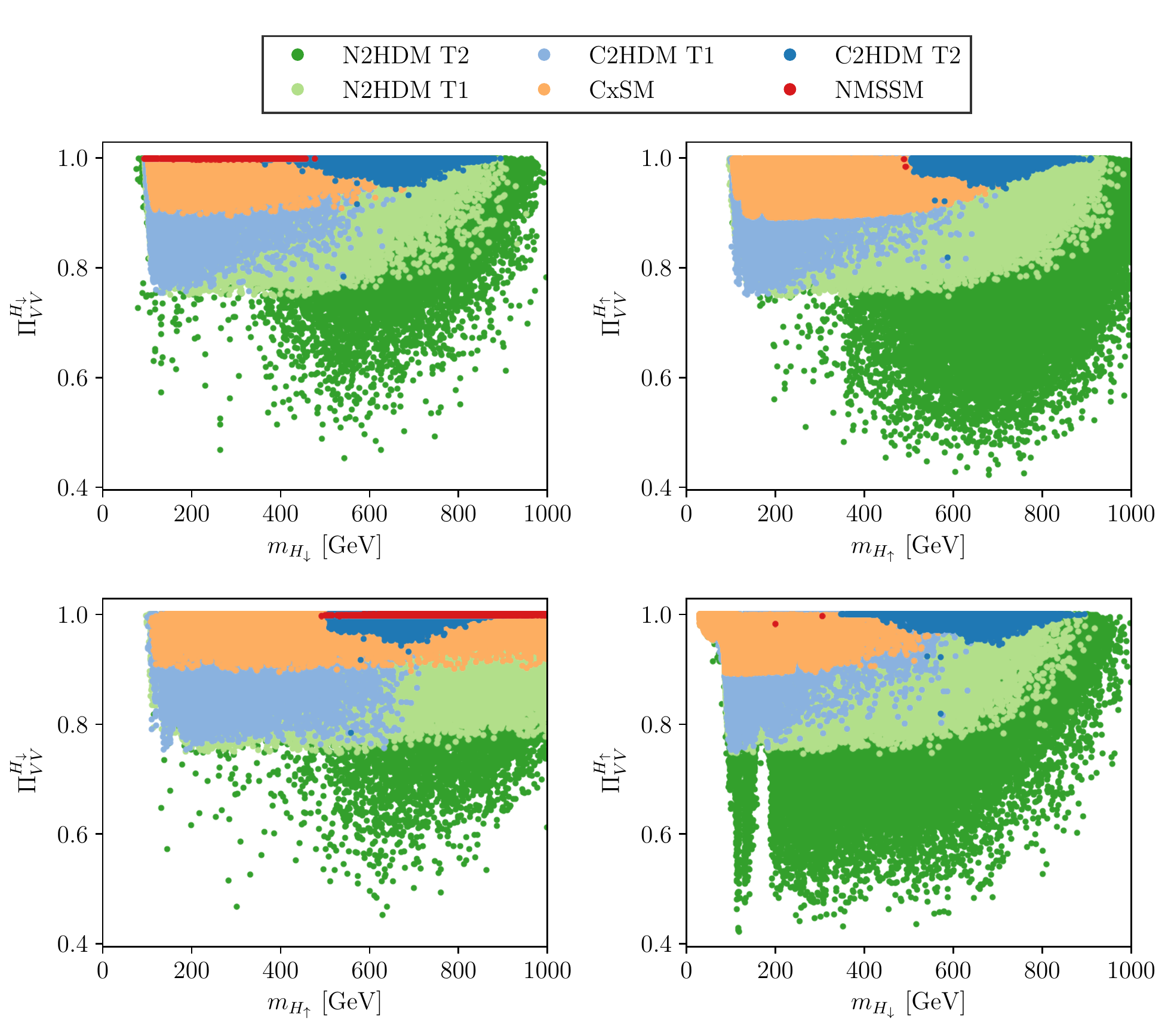}
\caption{The partial gauge boson sum $\Pi^{(2)}_{VV}$ assuming the only
  additionally discovered Higgs boson is $H_\downarrow$ (left) or
  $H_\uparrow$ (right) as a function of their respective mass (upper)
  and as a function of the mass of the non-discovered Higgs boson,
  respectively, (lower), for the CxSM (yellow), the type I N2HDM (fair
  green) and C2DHM (fair blue), the type II N2HDM (dark green) and
  C2HDM (dark blue) and the NMSSM (red).}\label{fig:sumrule}
\end{figure}

For all of our models we have
\beq
\Pi^{(3)}_{VV} = 1 \quad \mbox{ for the CxSM, N2HDM, NMSSM, C2HDM}\;,
\label{eq:vvsumrule}
\eeq
whereas in models with smaller Higgs sectors as the CP-conserving 2HDM
or the MSSM, the gauge boson sum reads
\beq
\Pi^{(2)}_{VV} = 1 \quad \mbox{ for the MSSM and the CP-conserving 2HDM}\;.
\eeq
Figure~\ref{fig:sumrule} shows the distribution of the partial gauge
boson sum $\Pi^{(2)}_{VV}$ for our models. We assume that besides
$h_{125}$ only one additional CP-even (or, for the C2HDM, CP-mixed) Higgs boson has
been discovered. In this case, the sum rule Eq.~(\ref{eq:vvsumrule})
is necessarily violated, as we only sum over two instead of three
Higgs bosons, and we expect to see deviations of $\Pi^{(2)}_{VV}$ from 1.
In the left column, we assume that the additionally discovered Higgs
boson is the $H_\downarrow$, and in the right one, it is assumed to
be the $H_\uparrow$. Without the
discovery of the third Higgs boson, we cannot decide in which of the
two situations we are. The upper (lower) row shows the distributions as a
function of the (non-)discovered Higgs boson mass, respectively.
All the points that are shown respect all our constraints,
including the requirement of Eq.~(\ref{eq:cpnature}).
We immediately observe, that $\Pi^{(2)}_{VV}$ cannot drop below about 0.9
  in the CxSM. This is a consequence of the simple coupling structure
  combined with the bound from the global signal strength, enforcing
  $c^2 (h_{125} VV) \gsim 0.9$ or equivalently $\Pi^{(2)}_{VV} \gsim 0.9$,
  even if the discovered non-SM Higgs does not couple to
  $VV$. Hence, deviations by more than 0.1 from the total gauge boson sum would
  allow to exclude the CxSM, although it is more likely that the
  CxSM can be excluded by deviations from the common coupling scaling,
  before the coupling sum analysis can be performed. \s

In the C2HDM type II, apart from very few outliers\footnote{These reside
in the wrong-sign regime yet not in the limit of the real 2HDM.},
the $h_{125}$ coupling squared to massive gauge bosons can
deviate by at most 10\% from the squared SM-value, {\it
  cf.}~Fig.~\ref{fig:C2HDM-coup}, which is reflected in the
outcome of the gauge coupling sum shown here.\footnote{The larger
  deviations in Fig.~\ref{fig:C2HDM-coup}, beyond 10\%, are in the
  limit of the real 2HDM. In this
  case, however, the gauge boson sum is saturated and we have
  $\Pi^{(2)}_{VV}=1$.} In the C2HDM type I on the other
hand larger deviations from the SM-limit are still possible, {\it
  cf.}~Fig.~\ref{fig:t1C2HDM-coup}, so that
the partial gauge coupling sum can become as small as 0.73. \s

In the N2HDM type II (type I) deviations from 1 of up to 55\% (25\%)
in $c^2(h_{125} VV)$ are possible, inducing the largest deviations of
all models from $\Pi_{VV}=1$. They are also larger than those attained
by the outliers in C2HDM type II. Moreover, the few outliers in
the C2HDM that can reach a violation of 35\% are likely to be probed
before a coupling sum analysis can be performed. The NMSSM, on the
other hand, although featuring the largest Higgs sector, is the most
constrained of our models because of supersymmetric relations.  As a
consequence, the coupling sum can deviate by at most a few percent if
the second discovered Higgs boson is $H_\downarrow$. In case it is the
heavier one, we hardly have any points that fulfil
  the requirement of rates above 10~fb in the $Z$ boson final state,
  {\it cf.}~Fig.~\ref{fig:Hd-ZZ}. In this case, the coupling sum deviates a bit more
from 1, by up to about 5\%. \s

In summary, the answers to our questions are, that all models
feature points where the gauge coupling sum is very close to 1 or
equal to 1 making
it very hard to distinguish them from the real 2HDM or the MSSM. This
is not surprising as all our models contain the alignment limit, {\it
  cf.}~also \cite{Haber:2017udj}. On the other hand, in
all our models there exist parameter configurations (although very
rare for the NMSSM and the C2HDM type II) where considerable deviations
from 1 allow for an easy discrimination from the Higgs sectors with two neutral Higgs
bosons. The coupling sum analysis allows for the exclusion of the CxSM if
$\Pi^{(2)}_{VV}$ deviates by more than 10\% from 1,
while in the C2HDM it would
indicate the realization of the wrong-sign
regime. As the lower plots reveal, a correlation between the
pattern of the coupling sum and the mass
scale of the escaped Higgs boson cannot be observed after taking into
account all mentioned constraints.
Finally, the observation of deviations by more than 35\%
singles out the N2HDM as a possible underlying model.

\subsection{Yukawa Coupling Sums}
\begin{figure}[t!]
  \centering
  \includegraphics[width=0.8\linewidth]{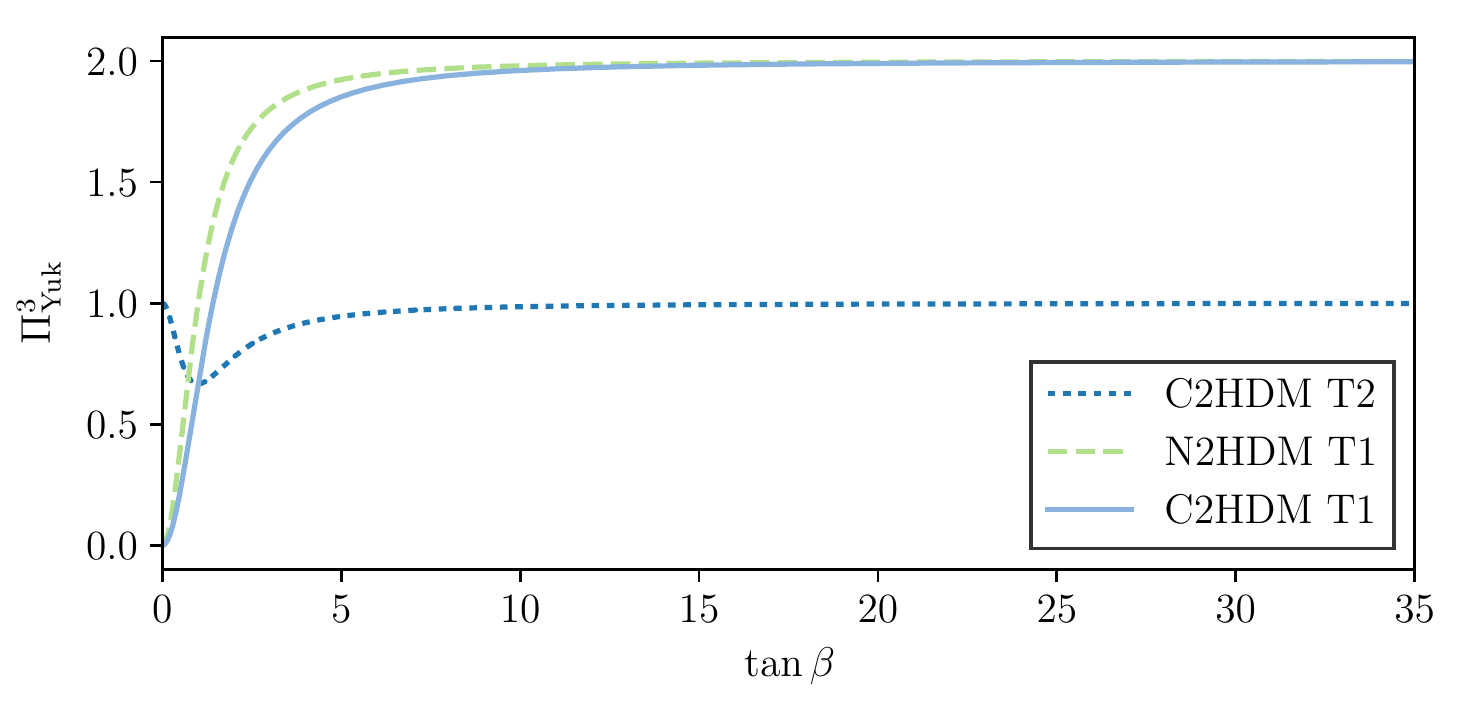}
  \caption{The non-trivial Yukawa sums for the C2HDM type II (short-dashed blue),
    Eq.~(\ref{eq:sumyukc2hdm}), the N2HDM type I (dashed
    green), Eq.~(\ref{eq:sumyukn2hdm1}) and the C2HDM type I (full
    blue), Eq.~(\ref{eq:sumyukc2hdm1}),  as a function of
    $\tan\beta$.}\label{fig:pi3tanbeta}
\end{figure}

The CxSM fulfils the Yukawa coupling sum
\beq
\Pi^{(3)}_{\text{Yuk}} = 2 \;.
\eeq
In the NMSSM and in the type II (as well as the lepton-specific) N2HDM we
have
\beq
\Pi^{(3)}_{\text{Yuk}} = 1 \;.
\eeq
The flipped N2HDM implies the same coupling sum if the $\tau$-lepton
coupling is exchanged by the $b$-quark coupling.
For the C2HDM Yukawa sum we use the effective fermion coupling squared
$|c(Hf\bar{f})|^2\equiv(c^e)^2+(c^o)^2$, with $c^e$ and $c^o$ defined in
Eq.~(\ref{eq:yuklag}).
For the C2HDM type II this leads to the sum
\beq
\Pi^{(3)}_{\text{Yuk}} = 2 \left( \frac{24}{17 - \cos(4\beta)} - 1 \right)
\;. \label{eq:sumyukc2hdm}
\eeq
The Yukawa sum as a function of $\tan\beta$ is shown in
Fig.~\ref{fig:pi3tanbeta} (short-dashed blue line). It has a minimum of
$\Pi^{(3)}_{\text{Yuk}}=2/3$ at $\tan\beta =1$ and quickly approaches 1
from below for all other $\tan\beta$ values.  \s

In type I, the Yukawa sum is the same both for the
$\tau$-lepton and the $b$-quark choice of the down-type fermion
coupling. In the N2HDM type I, we have
\beq
\Pi^{(3)}_{\text{Yuk}} = 2 \sin^2 \beta \;,\label{eq:sumyukn2hdm1}
\eeq
 and in the C2HDM type I the sum reads
\beq
\Pi^{(3)}_{\text{Yuk}} = \frac{2\tan^2\beta}{2+\tan^2\beta}.\label{eq:sumyukc2hdm1}
\eeq
Both of these are also shown in Fig.~\ref{fig:pi3tanbeta} (dashed green and full blue line, respectively). Flavour constraints~\cite{Mahmoudi:2009zx,Hermann:2012fc} require $\tan\beta\ge2.2$ in type I models which means that $\Pi^{(3)}_\mathrm{Yuk}$ cannot be much smaller than 2 ({\it cf.}~Fig.~\ref{fig:pi3tanbeta}) in both the C2HDM and N2HDM type I.
The result for the sums is the same in the flipped type, and
also in the lepton-specific case if
the $b$-quark is used instead of the $\tau$-lepton. For the real 2HDM
(MSSM) the Yukawa sums are the same as in the N2HDM
(NMSSM) with the difference that one only sums over 2 instead of 3
neutral Higgs bosons. \s

\begin{figure}[t!]
\centering
\includegraphics[width=0.9\linewidth]{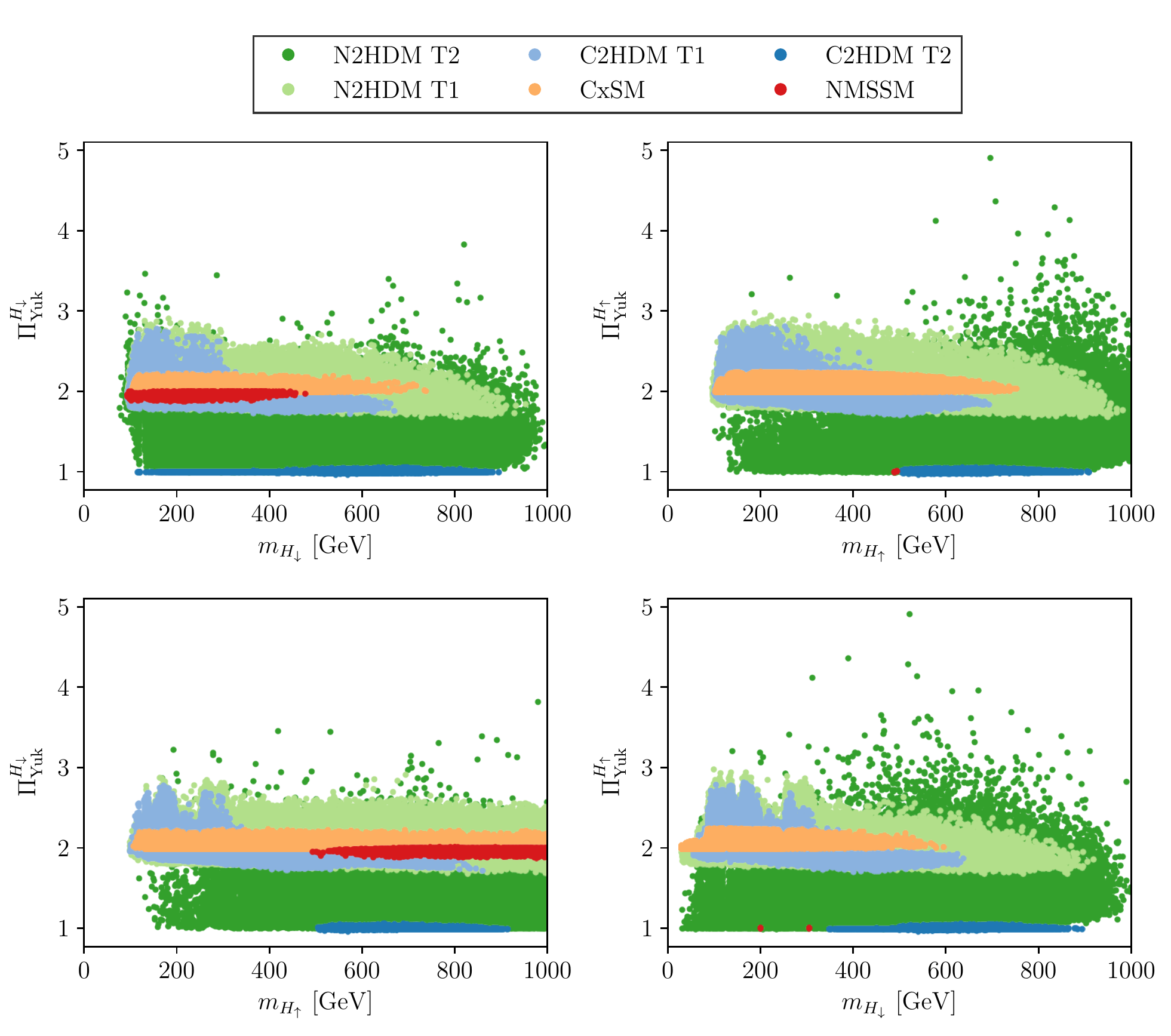}
\caption{Same as Fig.~\ref{fig:sumrule} but for the partial Yukawa sum
  $\Pi^{(2)}_{\text{Yuk}}$.}\label{fig:yuksumrule}
\end{figure}

In Fig.~\ref{fig:yuksumrule} the distributions of the partial Yukawa sums
$\Pi^{(2)}_{\rm Yuk}$ are depicted for the two situations where the
additionally discovered CP-even (or, for the C2HDM, CP-mixed) Higgs
boson is either the
lighter (left column) or the heavier (right column) of the non-SM Higgs bosons.
The upper (lower) row again shows the distributions as a
function of the (non-)discovered Higgs boson mass, respectively. \s

We observe, that due to the common rescaling of all CxSM Higgs couplings the lower
bound of the partial Yukawa sum is given by 2 with the maximal
violation of the complete sum given by the bound of the global signal
strength. The measurement of a value below 2 would
immediately exclude the CxSM. A measurement of $\Pi^{(2)}_{\text{Yuk}} <1$
on the other hand, would be very interesting because it is only
possible in the C2HDM type II and due to the
specific pseudoscalar admixture to the Yukawa couplings. Therefore, not only
the model but also the structure of the Yukawa sector could be
determined. According to Eq.~(\ref{eq:sumyukc2hdm}) it would also fix
$\tan\beta \approx 1$. Since the deviations from 1 can at most be
a few percent, however, the model is most probably
identified earlier through
other observables. The C2HDM type II is ruled out if deviations
larger than $7$\% above 1 are measured. In the C2HDM type I the values of the partial
Yukawa sum are distributed between about 1.7 and 2.8. The lower limit
is due to the lower bound on $\tan\beta$ imposed by the flavour
constraints. The observation of any violation below 1.7 and above
about 2.8 immediately excludes the C2HDM type I. This also applies for the
N2HDM type I where the maximum deviations range between the partial
sum values 1.7 and 2.8. \s

In the NMSSM the partial Yukawa sum is strongly violated with values between
1.8 and 2 if
the additional discovered Higgs boson is
the lighter one. If instead $H_\uparrow$ is discovered the Yukawa sum
is close to the saturating value of 1.
These two very different violation patterns allow to decide which of the two
non-SM-like Higgs bosons has been discovered, if one is able to
identify the NMSSM as the underlying model. The NMSSM is excluded if
violations beyond 2 are discovered. The large violations in case
$H_\downarrow$ is discovered can be explained by the fact that the
constraints applied on the NMSSM restrict $\tan\beta$ to small
values. For these, however, the heavier CP-even Higgs
boson $H_\uparrow$ is dominantly doublet-like, {\it cf.}~Table~2 in
\cite{King:2014xwa}. While $h_{125}$
carries most of the top-Yukawa coupling to comply with the Higgs data,
the non-discovered doublet-like $H_\uparrow$ has a large
coupling component to the down-type fermions. Its non-discovery leads
to the observed large violations in $\Pi^{(2)}_{\text{Yuk}}$. The
situation is reversed if $H_\uparrow$ is discovered. The $H_\downarrow$
is mostly singlet-like and its non-discovery barely violates the
Yukawa sum, which is close to 1. Finally, the
N2HDM type II with its large number of parameters not restricted
by SUSY relations can violate the Yukawa sum by a factor of almost
5. Any measurement of $\Pi^{(2)}_{\text{Yuk}}$ beyond about 2.9 therefore
clearly singles out the N2HDM type II among our candidate models. \s

In summary, the answers to our questions are: The type I C2HDM and
N2HDM, the type II N2HDM, the CxSM and the NMSSM all feature points
around the value $2\sin^2\beta$, that is obtained for the sum of
the 2HDM type I, so that a distinction from this model would then not
be possible. However, if the two discovered Higgs bosons are those of
the type I C2HDM or N2HDM, the CxSM or the two lighter Higgs bosons of
the NMSSM, then their sum would clearly exclude the possibility
of the 2DHM type II or the MSSM, as these lead to the sum value
of 1. The scale of the non-discovered Higgs boson cannot be
determined from the pattern of the Yukawa coupling sums. Only in the
NMSSM coupling sums close to 1 would indicate that the discovered
Higgs boson is the $H_\uparrow$, and above 1.8, that it is the
$H_\downarrow$. The distinction of the models, or at
least the exclusion of some of the models is possible as described in
the previous paragraph.

\begin{figure}[t!]
\centering
\includegraphics[width=0.9\linewidth]{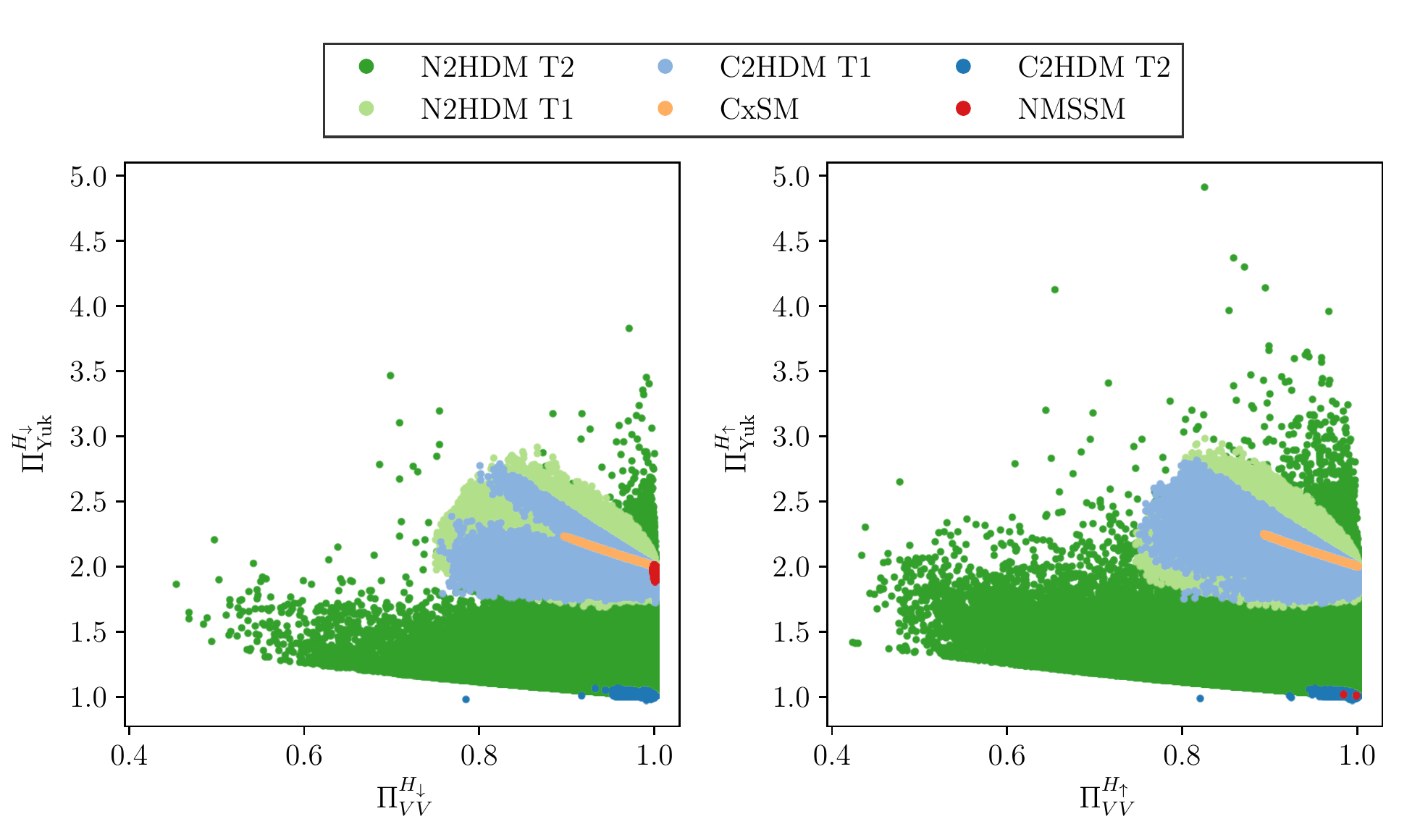}
\caption{Partial Yukawa sum $\Pi^{(2)}_{\text{Yuk}}$ versus $\Pi^{(2)}_{VV}$ in case
  $H_\downarrow$ (left) or $H_\uparrow$ (right) has been discovered.}\label{fig:corrsumrule}
\end{figure}
\subsection{Coupling Sum Correlations}
The previous discussions have already made clear that there are
correlations between the gauge boson and Yukawa coupling sums
that may be exploited. In Fig.~\ref{fig:corrsumrule} the partial sums
are plotted against each other for all of our models and the two
different discovery situations. The CxSM shows the simplest behaviour where
the two sums are strongly correlated due to the common rescaling of
the couplings. As also all other models except for the C2HDM type II
cover (part of) this region this behaviour does not
allow to distinguish the
models. Only deviations from this correlation rule out the CxSM. In the NMSSM the
plot clearly shows the two distinct regions resulting from the discovery of either the $H_\downarrow$ or the $H_\uparrow$. However, it is impossible in both regions to distinguish the NMSSM from the other models using only these coupling sums. The N2HDM is by far
the least constrained of our models. It shows a sharp lower
boundary which is a result of the orthogonality of the mixing matrix
and not due to the physical constraints. Observing
$\Pi^{(2)}_{VV} < 1$ and $\Pi^{(2)}_{\text{Yuk}} \approx 1$ therefore
excludes all models with a
$3 \times 3$ mixing of the CP-even scalars. The other models do not have
any points in this region because of their specific Yukawa structure
and/or the influence of other constraints. The only model in our study
where such a situation could be realized is the C2HDM type II,
identifying it as the candidate underlying model in this case.

\section{Conclusions \label{sec:concl}}
In this paper we investigated extensions of the SM, that
are motivated by specific features. Namely, they may solve some of the
problems of the SM, they are rather simple, and they feature 3 CP-even
Higgs bosons that have a singlet admixture. These are the CxSM, the
N2HDM and the NMSSM.
Additionally, we included the C2HDM as it also provides 3 neutral Higgs
bosons, which, however, now have a pseudoscalar admixture. This allows
us to compare the phenomenological implications of the different
admixtures. Furthermore, all these models are based on different
underlying symmetries that, again, are reflected in the phenomenology
of their Higgs bosons. In view of the non-discovery of new non-SM
particles, we investigated what can be learnt from the Higgs sector
itself. Our main focus was on the experimental situation where besides
the discovered SM-like Higgs boson only one additional Higgs is
discovered in a first stage. We considered the question: Can the
different models be distinguished based on the mass distributions, the
discovery rates as well as the gauge boson and Yukawa coupling sums?
Independently of this goal, the rates of all
neutral Higgs bosons for the investigated models in the various SM final
states presented here can be used as basis for further investigations, like {\it
  e.g.}~the identification of benchmark points. Note that all
generated parameter points fulfil the experimental and theoretical
constraints on the models. Our main
findings are the following: \s

The {\it EDM constraints}, that are relevant for the CP-violating 2HDM,
turn out to be more constraining in the C2HDM type II than in type I.
For a non-negligible CP-violating
phase, the Higgs mass spectrum is characterized by rather heavy non-SM-like Higgs bosons, with
both masses above about 500 GeV and not too far apart.\s

While either of the two lighter CP-even (CP-mixed in case of the
C2HDM) Higgs bosons can be the SM-like Higgs
in our models, the {\it mass spectrum} of the NMSSM and C2HDM type II
does not feature the possibility of the heaviest Higgs boson to be the
SM-like one. \s

We found that the present constraints all allow for a non-vanishing
singlet or pseudoscalar {\it admixture} to the $h_{125}$ that, depending on
the model, is more or less important and can be constrained by future
measurements of the rates. The results are summarized in
table~\ref{tab:admixtures}. The N2HDM results are taken from
  \cite{Muhlleitner:2016mzt}. Note that the upper bound on $\Psi$ for
  the C2HDM type II is mainly due to the EDM constraints. We also
  found that the C2HDM type II is the only model where
  $\mu_{\tau\tau}$ increases with rising value of the admixture. \s
\begin{table}[t!]
\begin{center}
\begin{tabular}{lcccccc} \toprule
Model & CxSM & C2HDM II & C2HDM I & N2HDM II & N2HDM I & NMSSM \\ \midrule
$\left(\Sigma\, {\rm or} \,\Psi\right)_{\text{allowed}}$ & 11\% & 10\% & 20\% & 55\% & 25\% &
 41\% \\
$\mu_{x} (\left(\Sigma\,{\rm or}\,\Psi\right)_{\text{max}})$ & global & all (7.5\%)  & $\mu_{VV}$
  (7\% ) & $\mu_{\tau\tau}$ (20-37\%) & $\mu_{VV}$ (7.5\%) &
                                                             $\mu_{\tau\tau}$ (8\%) \\
\bottomrule
\end{tabular}
\caption{2nd row: allowed singlet and pseudoscalar (for the C2HDM)
  admixtures; 3rd row: the most constraining $\mu_{xx}$ together with the
  maximum allowed admixture after a measurement of $\mu_{xx}$ within
  5\% of the SM value. The allowed CxSM
  admixture scales with the global signal strength. The first (second) value
  for N2HDM II applies for medium (small) $\tan\beta$ values. \label{tab:admixtures}}
\end{center}
\end{table}

From the investigation of the {\it rates} of the non-SM-like Higgs bosons,
which can be CP-even, CP-odd or CP-mixed states we concluded, regarding the observation of one additional neutral Higgs boson $\Phi$ besides the $h_{125}$, that:
\begin{itemize}
\item The CxSM is excluded in any of the SM final state channels if
  rates above the SM reference are found.
\item In the $ZZ$ final state, rates of ${\cal O}(100)$~fb for $m_\Phi
  \gsim 380$~GeV exclude the NMSSM, and rates of
  35-50~fb for $m_\Phi \in [800 : 1000]$~GeV are only possible in the
  N2HDM II. The latter rates are about to be probed
  by the
  experiments, however.
\item In the $\tau\tau$ final state, we find strongly enhanced rates for
  the C2HDM and N2HDM type II in the limit of the real 2HDM for $m_\Phi \in
  \left[70 : 80\right]$~GeV. This is also the case for the pseudoscalar decay in
  the N2HDM type II. This should encourage the experiments to extend their
  analyses to this mass range. Overall, no distinction of the models is
  possible based on this final state rate alone.
\item In the $\gamma\gamma$ final state, we find again strongly enhanced rates
  for $m_\Phi \in[ 70 : 80]$~GeV in the C2HDM for $\Psi \to 0$,
as the experiments have not provided
  exclusion limits here yet. Furthermore, rates above 5~fb for $m_\Phi \in
  [130 : 350]$~GeV single out the N2HDM II as the only allowed model in our set.
\item In the $t\bar{t}$ final state the N2HDM, C2HDM and NMSSM rates
  can be above the SM reference and even reach ${\cal O}(100)$~pb
  in the N2HDM rendering the search for the
  additional Higgs bosons in this final state
  interesting. Note that the NMSSM is found to be excluded if rates above
  about 4~pb are measured.
\end{itemize}

The requirement of unitarity implies {\it coupling sum rules}. In case
not all the Higgs bosons that carry an electroweak doublet component are
found, these sum rules are violated. This gives a handle to decide
whether the discovered Higgs spectrum is complete or not. Thus, we
investigated the partial gauge boson and Yukawa sums assuming
that only one additional Higgs boson besides the $h_{125}$ has been
discovered. In our models with three CP-even (CP-mixed, for the C2HDM)
Higgs bosons this inevitably induces violations of the coupling sums. We
found, for all our models, that the partial gauge boson sum contains
points where the sum rule is fulfilled. This is to be expected, as the
$h_{125}$ couples almost SM-like to the massive gauge bosons. In this
case, a distinction from the MSSM or the 2HDM with two CP-even Higgs
bosons is impossible. Also in the partial Yukawa sums we
found scenarios fulfilling the complete sum. There are a lot of
scenarios, however, that violate the complete coupling sums and that can be used to
identify some distinguishing features:
\begin{itemize}
\item The violation of the gauge boson sum rule by more than 10\% excludes
  the CxSM, the violation by more than 35\% singles out the N2HDM as
  a candidate underlying model. In case the NMSSM can be identified as the
  underlying model, by finding {\it e.g.} additional supersymmetric
  particles, the violation of the sum rule would allow it to be
  distinguished from the MSSM, for which the sum rule is saturated
  after the discovery of two CP-even Higgs bosons. Measurable
  violations of the gauge boson sum rule are, however, only observed
  if the additionally discovered Higgs bosons is the $H_\uparrow$ and
  if $H_\downarrow$ has a mass near 125~GeV. 
\item The violation of the Yukawa sum with values below 2 excludes the
  CxSM and with values above 2 the NMSSM. In case the partial Yukawa
  sum yields values below 1, the candidate model is the C2HDM II. The C2HDM II on the
  other hand is excluded for values above 1.07. If the partial sum
  yields values below 1.7 or above 2.85 then the type I C2HDM or N2HDM
  are excluded. If it is a supersymmetric model and deviations by more
  than 80\% away from 1 are observed, then the
  candidate is the NMSSM and, in that case, the
  $H_\downarrow$ has been discovered. For values close to 1 it is the
  $H_\uparrow$ that has been discovered. Finally, values above 2.9
for the partial Yukawa sum
  single out the N2HDM II as the candidate underlying model.
\end{itemize}

Our results show, that even if only a subset of the Higgs bosons of
extended Higgs sectors is found, the use of the Higgs rates and
coupling sums and their combination may allow for the distinction of
models and eventually even for the identification of a specific candidate
model. The next step to be taken now is the definition of benchmark
points that feature specific properties to support this task and that
the experiments can include in their experimental analyses.

\subsubsection*{Acknowledgements}
The authors acknowledge financial support from the DAAD
project ``PPP Portugal 2015'' (ID: 57128671) and from the FCT project ``Exploring Beyond the Standard Model Higgs Sectors''.
M.S. is funded through the
grant BPD/UI97/5528/2017. The work in this paper was also supported by
the CIDMA project UID/MAT/04106/2013.
We would like to thank
Philipp Basler, Rohini Godbole and Michael Spira for useful discussions.

\vspace*{1cm}
\bibliographystyle{h-physrev}

\end{document}